\documentclass[12pt]{article}

\usepackage{epsfig,epsf}

\begin{document}

\def\beq{\begin{equation}}

\def\eeq{\end{equation}}

\def\bea{\begin{eqnarray}}

\def\eea{\end{eqnarray}}

\def\eq#1{{Eq.~(\ref{#1})}}

\def\fig#1{{Fig.~\ref{#1}}}
             
\newcommand{\bas}{\bar{\alpha}_S}

\newcommand{\tk}{\tilde{\kappa}}
\newcommand{\as}{\alpha_S} 

\newcommand{\bra}[1]{\langle #1 |}

\newcommand{\ket}[1]{|#1\rangle}

\newcommand{\bracket}[2]{\langle #1|#2\rangle}

\newcommand{\intp}[1]{\int \frac{d^4 #1}{(2\pi)^4}}

\newcommand{\mn}{{\mu\nu}}

\newcommand{\tr}{{\rm tr}}

\newcommand{\Tr}{{\rm Tr}}

\newcommand{\T} {\mbox{T}}

\newcommand{\braket}[2]{\langle #1|#2\rangle}

\newcommand{\ab}{\bar{\alpha}_S}

\newcommand{\x}{\vec{x}}
\newcommand{\y}{\vec{y}}
\newcommand{\z}{\vec{z}}
\newcommand{\rr}{\vec{r}}
\newcommand{\bb}{\vec{b}}
\newcommand{\bbb}{\vec{b}^{\,'}}
\newcommand{\xx}{\vec{x}^{\,'}}
\newcommand{\yy}{\vec{y}^{\,'}}
\newcommand{\zz}{\vec{z}^{\,'}}
\newcommand{\rrr}{\vec{r}^{\,'}}

\setcounter{secnumdepth}{7}

\setcounter{tocdepth}{7}

\parskip=\itemsep               %?

\setlength{\itemsep}{0pt}       %?

\setlength{\partopsep}{0pt}     %?

\setlength{\topsep}{0pt}        %?

%---layout fuer eine dina4 seite-------------------

\setlength{\textheight}{22cm}

\setlength{\textwidth}{174mm}

\setlength{\topmargin}{-1.5cm}

%\input psfig

%%%%%%%%%%%%%%%%%%%%%%%%%%%%%%%%%%%%%%%%%%%%%%%%%%%%%%%%%%%%%%%

%\renewcommand{\thefootnote}{\fnsymbol{footnote}}

%\newcommand{\m}{\marginpar{*}}

\newcommand{\lash}[1]{\not\! #1 \,}

\newcommand{\pd}{\partial}
\newcommand{\h}{\frac{1}{2}}

\newcommand{\g}{{\rm g}}

\newcommand{\el}{{\cal L}}

\newcommand{\A}{{\cal A}}

\newcommand{\Ka}{{\cal K}}

\newcommand{\al}{\alpha}

\newcommand{\be}{\beta}

\newcommand{\ep}{\varepsilon}

\newcommand{\ga}{\gamma}

\newcommand{\de}{\delta}

\newcommand{\dt}{\delta^{(2)}}

\newcommand{\De}{\Delta}

\newcommand{\et}{\eta}

\newcommand{\ka}{\vec{\kappa}}

\newcommand{\la}{\lambda}

\newcommand{\ph}{\varphi}

\newcommand{\si}{\sigma}

\newcommand{\ro}{\varrho}

\newcommand{\Ga}{\Gamma} 

\newcommand{\om}{\omega}

\newcommand{\La}{\Lambda}  

\newcommand{\tG}{\tilde{G}}

\newcommand{\lb}{\left(}
\newcommand{\rb}{\right)}

\renewcommand{\theequation}{\thesection.\arabic{equation}}

%%%%%%%%%%%%%%%%%%%%%%%%%%%%%%%%%%%%%%%%%%%%%%%%%%%%%%%%%%%%%%%%%

% ABBREVIATED JOURNAL NAMES  

%

\def\ap#1#2#3{     {\it Ann. Phys. (NY) }{\bf #1} (19#2) #3}

\def\arnps#1#2#3{  {\it Ann. Rev. Nucl. Part. Sci. }{\bf #1} (19#2) #3}

\def\npb#1#2#3{    {\it Nucl. Phys. }{\bf B#1} (19#2) #3}

\def\plb#1#2#3{    {\it Phys. Lett. }{\bf B#1} (19#2) #3}

\def\prd#1#2#3{    {\it Phys. Rev. }{\bf D#1} (19#2) #3}

\def\prep#1#2#3{   {\it Phys. Rep. }{\bf #1} (19#2) #3}

\def\prl#1#2#3{    {\it Phys. Rev. Lett. }{\bf #1} (19#2) #3}

\def\ptp#1#2#3{    {\it Prog. Theor. Phys. }{\bf #1} (19#2) #3}

\def\rmp#1#2#3{    {\it Rev. Mod. Phys. }{\bf #1} (19#2) #3}

\def\zpc#1#2#3{    {\it Z. Phys. }{\bf C#1} (19#2) #3}

\def\mpla#1#2#3{   {\it Mod. Phys. Lett. }{\bf A#1} (19#2) #3}

\def\nc#1#2#3{     {\it Nuovo Cim. }{\bf #1} (19#2) #3}

\def\yf#1#2#3{     {\it Yad. Fiz. }{\bf #1} (19#2) #3}

\def\sjnp#1#2#3{   {\it Sov. J. Nucl. Phys. }{\bf #1} (19#2) #3}

\def\jetp#1#2#3{   {\it Sov. Phys. }{JETP }{\bf #1} (19#2) #3}

\def\jetpl#1#2#3{  {\it JETP Lett. }{\bf #1} (19#2) #3}

%%%%%%%%% notice the parenthesys is only on one side

\def\ppsjnp#1#2#3{ {\it (Sov. J. Nucl. Phys. }{\bf #1} (19#2) #3}

\def\ppjetp#1#2#3{ {\it (Sov. Phys. JETP }{\bf #1} (19#2) #3}

\def\ppjetpl#1#2#3{{\it (JETP Lett. }{\bf #1} (19#2) #3} 

\def\zetf#1#2#3{   {\it Zh. ETF }{\bf #1}(19#2) #3}

\def\cmp#1#2#3{    {\it Comm. Math. Phys. }{\bf #1} (19#2) #3}

\def\cpc#1#2#3{    {\it Comp. Phys. Commun. }{\bf #1} (19#2) #3}

\def\dis#1#2{      {\it Dissertation, }{\sf #1 } 19#2}

\def\dip#1#2#3{    {\it Diplomarbeit, }{\sf #1 #2} 19#3 }

\def\ib#1#2#3{     {\it ibid. }{\bf #1} (19#2) #3}

\def\jpg#1#2#3{        {\it J. Phys}. {\bf G#1}#2#3}  

%

%%%%%%%%%%%%%%%%%%%%%%%%%%%%%%%%%%%%%%%%%%%%%%%%%%%%%%%%%%%%%%%%%%%%%

%

%\renewcommand{\thefigure}{{\protect\bf\arabic{figure}}}

\def\thefootnote{\fnsymbol{footnote}} 

%

%

%

%\begin{titlepage}

\noindent

\begin{flushright}

\parbox[t]{10em}{ \tt{TAUP 2795-05}\\
{\tt \today} \\
{\tt hep-ph/0502243}
 }
\end{flushright}

\vspace{1cm}

\begin{center}

{\LARGE  \bf High energy amplitude in the dipole approach }\\
{\LARGE  \bf with Pomeron loops: asymptotic solution}\\

\vskip1cm

{\Large \bf   ~E. ~Levin ${}^{\ddagger}$
\footnotetext{${}^{\ddagger}$ \,\,Email:
leving@post.tau.ac.il, levin@mail.desy.de, elevin@quark.phy.bnl.gov } }

\vskip1cm

{\it \,\,\, HEP Department}\\
{\it School of Physics and Astronomy}\\
{\it Raymond and Beverly Sackler Faculty of Exact Science}\\
{\it Tel Aviv University, Tel Aviv, 69978, Israel}\\
\vskip0.3cm

\end{center}  

\bigskip

\begin{abstract}        

In this paper an  analytical solution for the high energy scattering amplitude is suggested.  
This solution has several unexpected features:(i) the asymptotic amplitude is a function of dipole
 sizes and, therefore, this  amplitude shows  the gray disc structure at high energy,  instead of 
black disc behaviour  which was expected; (ii) the amplitude approaches the asymptotic limit in the 
same way as the solution to the Balitsky-Kovchegov equation does  ($\propto\,\exp(- C Y^2) $ ), but 
the coefficient $C$ in eight times smaller than for  the Balitsky-Kovchegov equation; (iii) the 
process of merging of two dipoles into one, only   influences  the high energy asymptotic behaviour 
by changing the initial condition from $Z(Y; [u_i = 1]) = 1 $ to   $Z(Y; [u_i = 1 - \gamma_{0,i}]) 
=1$. The value of $\gamma_0$ is determined by the  process of merging of two dipoles into one.  
With  this new initial condition  the Balitsky-JIMWLK approach  describes the high energy  
asymptotic behaviour of the scattering amplitude without any modifications recently suggested.

\end{abstract}

\newpage

%*********************************************************************************  

%*********************************************************************************  

\def\thefootnote{\arabic{footnote}} 
\section{Introduction}

Our approach to high energy interactions  in QCD is based
\cite{GLR,MUQI} on the BFKL Pomeron \cite{BFKL} and on the reggeon-like diagram technique
which takes into account the interactions of  BFKL Pomerons \cite{BART,NP,BLV}. It is well
known  \cite{GRPO,LEPO,BOPO} that the Pomeron diagrams technique can  be
re-written as a Markov processes \cite{GARD} for the  probability of finding  a given number of
Pomerons at fixed rapidity $Y$.

 However,
the colour dipole model  provides an alternative approach \cite{MUCD}, in which we can
replace the non-physical probability to
find several Pomerons\footnote{We call this probability non-physical since we cannot suggest an
experiment in which we could measure this observable.}, by the probability to find a given
number of dipoles. It has been shown that everything  we had know   about high energy
interactions can  be re-written in the leading $N_c$ approximation ($N_c$ is the number of
colours),in probabilistic approach\cite{MUCD,LL,LLB}  using  dipoles and their interactions.
 In  this approach assuming that we neglect the correlations inside the target, we have a non-linear 
evolution
equation (Balitsky-Kovchegov equation \cite{B,K}, see  Ref. \cite{LL,JANIK} for  alternative 
attempts).

The alternative approaches  based on the idea of strong gluonic fields
\cite{MV},  or on the Wilson loops approach \cite{B},  lead to a different type  of theory which is,
 at first sight,  not related to  Pomeron interactions.  For example the JIMWLK equation 
\cite{JIMWLK}
appears    to be      quite different to the Balitsky-Kovchegov equation. However, it has been shown
that the  Balitsky- JIMWLK approach,  is the same as the
method  based on the dipole model in the leading $N_c$ approximation (see Ref.
\cite{KOLU1,KOLU2}).  The further development of the Balitsky - JIMWLK formalism, as well as a deeper
understanding  the interrelations of this formalism with the old reggeon - like technique based on
the
BFKL Pomerons and their interactions,  is under  investigation 
\cite{KOLU2,KOLU3,BIT} and we  hope for  more progress in this direction.
Although,  the Balitsky-Kovchegov equation is widely used for  phenomenology,  it has a very
restricted region of applicability \cite{IM,KOLE,MUSO},  since it has been derived assuming
that only the process of the  decay of one dipole into two dipoles is essential at high energy.
As was shown by Iancu and Mueller \cite{IM} (see also Refs.
\cite{KOLE,MUSO,IT1,MSW,LLTR,IT2}),  a theory neglecting  the merging of two dipoles, is not
 only incorrect,  but that the corrections
from the process of annihilation of two dipoles into one dipole  should be taken into
account, to
satisfy  the unitarity constraint in the $t$-channel of the processes. This,  in spite of the fact 
that
there is  a $1/N^2_c$ suppression in the framework of the dipole approach.

A number of papers has been  circulated  recently  \cite{IT1,MSW,LLTR,IT2,KOLU1} in which the
key problem of   how to include the annihilation of dipoles without losing the
probabilistic interpretation of the approach is discussed.

In this paper we find the asymptotic solution to the high energy amplitude using  the approach
of Ref. \cite{LLTR}. The paper is organized as follows. In the next introductory  section  we
discuss
the dipole approach in $1/N^2_c$ approximation, and consider the main equations which we wish to
solve in this paper. We also discuss the statistical interpretation of these equations which 
play a major role in our method  of searching for the solution. The section 3 is devoted to solving 
the
simplified toy model where  the interaction does not depend on the size of interacting dipoles.
We  find here  that the large parameters of the problem allow us to use the semi-classical approach
for analyzing  the solution. In section 4 we find the asymptotic behaviour of the high energy
amplitude in the dipole approach including  Pomeron loops. We discuss how the amplitude  high energy
approaches this solution. In the short last section we summarize our results.

\section{Generating functional and statistical approach}
\begin{figure}[ht]
    \begin{center}
        \includegraphics[width=0.90\textwidth]{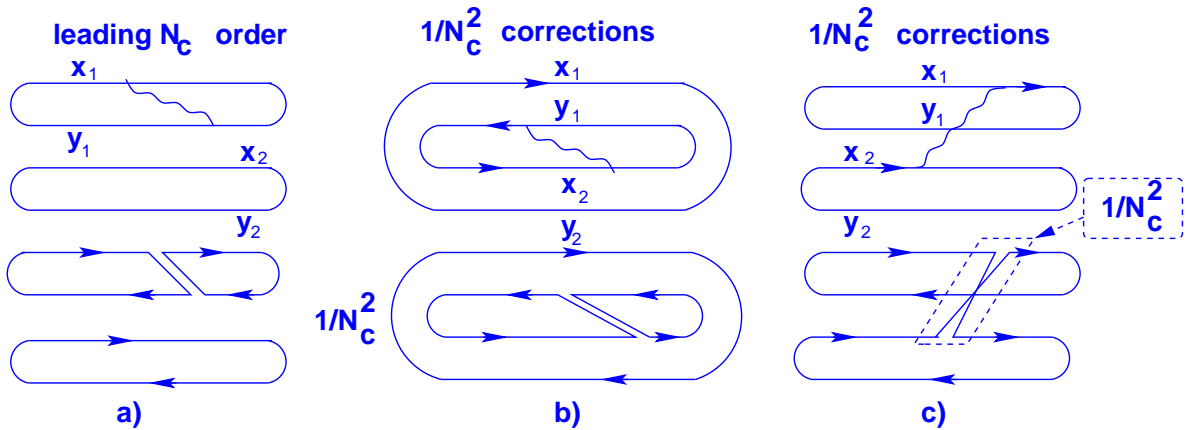}
        \caption{ \it Two dipole rescattering in leading $N_c$ order  (\fig{nccor}-a) and 
$1/N^2_c$
corrections (\fig{nccor}-b and \fig{nccor}-c).\fig{nccor}-b shows the  $1/N^2_c$ corrections due to 
re-grouping of the dipoles. This contribution has the same topology as the leading $N_c$ order 
diagram (see \fig{nccor}-a. \fig{nccor}-c corresponds to interaction of two quarks from the different 
dipoles. This contribution has a different topology than \fig{nccor}-a, and has to be assigned to so 
called multi-reggeon Pomerons (see for example Ref. \cite{KKM}.   }
 \label{nccor}
    \end{center}
\end{figure}

\begin{boldmath}
\subsection{Dipole model in $1/N^2_c$ approximation}
\end{boldmath}
The first question that we need to consider is, can we use the dipole approach for  calculating   
$1/N^2_c$ correction. At first sight the answer is  negative . Indeed, Mueller and Chen showed 
in Ref. \cite{MUCH} that the term of  order  $\bas/N^2_c$ cannot be rewritten as dipole 
interaction (see Ref. \cite{KOLU2}, in which this result is demonstrated  in the framework of 
the JIMWLK approach). We consider  two  dipole  rescattering given by \fig{nccor}.
In the leading order in $N_c$,  the gluon can be emitted by a  quark (or antiquark) of one dipole, 
and 
should be 
absorbed by an antiquark (quark) of the same dipole,  as  is shown in \fig{nccor}-a. In the next to 
the 
leading order in $N_c$ the interaction of two dipoles has the form \cite{MUCH,KOLU2}:
\beq \label{2D1}
\frac{\bas}{N^2_c}\,\int \,d^2 z\,K \lb \x_1,\y_1;\x_2,\y_2;\z \rb\,\,\langle T(\x_1,\y_1\,x_2,\y_2) 
\rangle
\eeq
where $T(\x_1,\y_1\,x_2,\y_2)$ is the interaction operator for two dipoles, $\bas = N_c \as/\pi$  and
\beq \label{2DK}
K \lb \x_1,\y_1;\x_2,\y_2;\z \rb\,\,=\,\,\lb \frac{(\x_1 - \z)_i}{(\x_1 - \z)^2}\,\,-\,\,\frac{(\y_1 
- \z)_i}{(\y_1 - \z)^2}\rb \,\times \,\lb \frac{(\x_2 - \z)_i}{(\x_2 - \z)^2}\,\,-\,\,\frac{(\y_2
- \z)_i}{(\y_2 - \z)^2}\rb \,=
\eeq
\beq \label{2DK1}
=\,\, \h\,\lb K(\x_2,\y_1;\z)\,\,+\,\, 
K(\x_1,\y_2;\z)\,\,-\,\,K(\x_1,\x_2;\z)\,\,-\,\,K(\y_1,\y_2;\z) \rb
\eeq
where
\beq \label{K}
K(\x,\y;\z)\,\,\,=\,\,\frac{(\x\,-\,\y)^2}{(\x\,-\,\z)^2\,\,(\y\,-\,\z)^2}
\eeq

\eq{2DK1}   has been discussed in Ref. \cite{HIIM} in the context of the odderon structure in 
the JIMWLK -approach. The first two terms in \eq{2DK1} describe the emission of a gluon by two dipoles 
$x_2 - y_1$ and $ x_1 - y_2$ (see \fig{nccor}-b). The configuration in which quark $ x_1$ (antiquark 
$y_1$)  and 
antiquark $y_2$ (quark $x_2$) creates a colorless pair is certainly suppressed by factor $1/N^2_c$.
However, after being created dipoles $x_2 - y_1$ and $ x_1 - y_2$,  will interact as two dipoles in 
the 
leading $N_c$ approximation (compare \fig{nccor} -a   and \fig{nccor}-b). Indeed, the topology of 
this term 
is the same as two cylinders \footnote{
The diagrams  describing a dipole target interaction in the leading order in $N_c$ has a cylinder 
topology as was shown in Ref. \cite{VE}. This  means that all diagrams for the BFKL Pomeron exchange  
can be drawn in the cylindric surface.}. Therefore for the first two terms we can replace
$\langle T(\x_1,\y_1\,x_2,\y_2) \rangle $ by the product $\langle T(\x_1,y_2) \rangle \,\,
\langle\,T(\x_2,\y_1) \rangle$.

The second two terms in \eq{2DK1} stem from the possibility of two quarks rescattering with a 
suppression of $1/N^2_c$. This interaction leads to a quite different topology (see \fig{nccor}-c), 
which cannot be treated as two independent parton showers. This new configuration with non-cylindric 
topology should be treated separately using the  so called BKP equation \cite{BKP},  and it is called  
multi-reggeon Pomeron. Such Pomerons have been studied  long ago (see Refs. \cite{LI,KKM}) and to 
the best of our knowledge, the intercepts of these $n$-reggeon Pomerons turn out to be smaller than 
the intercept of $n$ cylindrical  configurations (in our case $n =2$).

Therefore, we conclude that we can use the dipole model to even calculate  $1/N^2_c$ corrections.
This analysis is supported by direct calculation in Ref. \cite{BLV},  where it is shown that the 
triple BFKL Pomeron vertex calculated by Bartels in Ref. \cite{BART} generates $1/N^2_c$ corrections 
which can be rewritten as dipole interactions.

\subsection{Main equations}

In this section we will discuss the main equations of Ref. \cite{LLTR} paying our attention to their 
statistical interpretations.

In Ref.\cite{MUCD} the  generating functional which  characterizes  the system of interacting dipoles
 was introduced, it is is defined as

\beq \label{Z}
Z\left(Y\,-\,Y_0;\,[u] \right)\,\,\equiv\,\,
\eeq
$$
\equiv\,\,\sum_{n=1}\,\int\,\,
P_n\left(Y\,-\,Y_0;\,x_1, y_1;\,x_2, y_2;\, \dots ; x_i, y_i; \dots ;x_n, y_n
 \right) \,\,
\prod^{n}_{i=1}\,u(x_i, y_i) \,d^2\,x_i\,d^2\,y_i
$$
where $u(x_i, y_i) \equiv u_i $ is an arbitrary function of $x_i$ and $y_i$.
$P_n$ is a probability density to find $n$ dipoles with the size $x_i - y_i$, and with impact 
parameter $(x_i + y_i)/2$. Directly from the physical meaning of $P_n$ and definition in \eq{Z} it  
follows 
that the functional (\eq{Z}) obeys the condition
\beq \label{ZIN1} 
Z\left(Y\,-\,Y_0;\,[u=1]\right)\,\,=\,\,1\,.
\eeq
The physical meaning of (\eq{ZIN1}) is that the sum over
all probabilities is one.

The functional $Z$ has a very direct analogy in the  statistical approach: i.e.  the characteristic ( 
generating 
) function $\phi(\mathbf{s})$ in Ref. \cite{GARD}. For $P_n$ we have a typical birth-death equation 
which can be written in the form:
$$
\frac{\partial \,P_n(\dots,x_i,y_i; \dots)}{\partial Y}\,\,\,=
$$
\begin{eqnarray}
\,\,\,\,\,\,\,\,\,\,\,\,\,\,\,\,\,\,&=&\,\,- \sum_{i} \  \Gamma_{1 \to 
2} 
\bigotimes \lb P_n (\dots,x_i,y_i; \dots)\,\,-\,\,\,P_{n-1}(\dots,x_i,y_i; \dots) \rb \label{P} \\
 & & \,\,- \sum_{i \neq k}\, \Gamma_{2 \to 1}\bigotimes \lb P_n(\dots,x_i,y_i; \dots)\,-\,  
P_{n+1}(\dots; x_i,y_i; \dots ; x_k,y_k; \dots) \rb  \nonumber \\
 & & \,\,- \sum_{i \neq k}\, \Gamma_{2 \to 3}\bigotimes  \lb P_n(\dots,x_i,y_i; \dots; x_k,y_k;\dots)
 \,-\, P_{n -1}(\dots,x_i,y_i; \dots; x_k,y_k;\dots)\rb \nonumber 
\end{eqnarray}
In \eq{P} $\Gamma_{1 \to 2}$ is the vertex for the process of decay of one dipole with size $x 
- y$,  into two dipoles  with sizes $x - z$ and $y - z$. This 
vertex is well known and it is equal to
\beq \label{V12}
 \Gamma_{1 \to 2}\,\,\,=\,\,\bas\,K(\x,\y;\z)
\eeq
The vertex for the process of transition of two dipoles with sizes $x_1 - y_1$ and $x_2 - y_2$ into
three dipoles with sizes $x_1 - y_2$,$ z - y_1$ and $x_2 - z$ has been discussed in Ref. \cite{LLTR} 
and has the form
\beq \label{V23} 
\Gamma_{2 \to 3}\,\,=\,\,\frac{\bas}{2\,N^2_c}\,\lb K(\x_2,\y_1;\z) \,+\,K( \x_1,\y_2;\z) \rb
\eeq
We will discuss  the $2 \to 1$ vertex we will discuss  below.

$\bigotimes$ denotes all necessary integrations (see more detailed form in Ref. \cite{LLTR}). One 
can see that each line in \eq{P} gives  a 
balance of the death of a particular dipole ( the first term in each line which has a minus sign), 
and 
of the birth of two or three dipoles ( the second term in each line  which gives a positive 
contribution . \eq{P} is a typical equation for the  Markov's process ( Markov's chain) \cite{GARD}.

The last term in \eq{P} has the following form:
\beq \label{LT}
\Gamma_{2 \to 3}\bigotimes  \lb P_n(\dots,x_i,y_i; \dots; x_k,y_k;\dots)
 \,-\, P_{n -1}(\dots,x_i,y_i; \dots; x_k,y_k;\dots)\rb\,\,=\,\,
\eeq
$$
=\,\,\int\,d^2 x_i\,d^2 x_k \,d^2 y_i\,d^2 y_k\,d^2 z\,\frac{\bas}{2\,N^2_c}\,\lb 
K(\vec{x}_i,\vec{y}_k;\vec{z}) \,+\, K(\vec{x}_k,\vec{y}_i;\vec{z}) 
\rb\,
$$
$$
  \lb P_n(\dots,x_i,y_i; \dots; x_k,y_k;\dots)
 \,-\, P_{n -1}(\dots,x_i,y_i; \dots; x_k,y_k;\dots)\rb
$$

Multiplying \eq{P} by the product $\prod^n_{i=1}\,u_i$
and integrating over all $x_i$ and $y_i$,  we obtain the
following linear equation for the generating functional:
\beq\label{ZEQ}
\frac{\partial \,Z\,[\,u\,]}{
\partial \,Y}\,\,= \,\,\chi\,[\,u\,]\,\,Z\,[\,u\,]
\eeq
with
\beq  \label{chi}
\chi[u]=
\eeq
\begin{eqnarray}
&=& -\,\int\,d^4\,q d^4 q_1\,d^4 q_2 \,\,\lb  \Gamma_{1\,\rightarrow \,2}\lb q \to q_1 + q_2 \rb\
\lb - u(q) \,+\,u(q_1) \,u(q_2)\,\rb \,\frac{\delta}{\delta u(q)}\,- \right. \label{VE12} \\
 & &\left. -  \Gamma_{2\,\rightarrow \,1}\lb  q_1 + q_2 \to q  \rb\,
\lb u(q_1)  \,u(q_2) \,-\, u(q) \rb \,\,\h \,\frac{\delta^2}{\delta u(q_1)\,\delta u(q_2)} \rb\,- 
\label{VE21} \\
 & & -\int\,\prod^2_{i=1} d^4 q_i\,\prod^3_{i=1} d^2 \,z\,\Gamma_{2 \to 3}\lb (x_1,y_1) + 
(x_2,y_2) \to
(x_1,y_2) + (z,y_1) + (x_2,z)  \rb \,\nonumber\\
 & & \lb u(x_1,y_2)\,-\,1\rb\,\lb u(x_2,y_1)\,\,-\,\,\,u(z,y_1)\,u(x_2,z)\,\rb\,\h 
\,\frac{\delta^2}{\delta u(q_1)\,\delta u(q_2)} \label{VE23}
\end{eqnarray}
where we denote $d^4 q_i = d^2 x_i\,d^2 y_i$.
The functional derivative with respect to $u(q)=u(x,y)$,  plays the  role 
of an  annihilation operator for a dipole of the size $r = x - y$,  at the impact 
parameter $b= \h(x + y)$. 
The multiplication by $u(x,y)$ corresponds to
a creation operator for this dipole. Recall that $d^4 q_i $ stands for $d^2 x_i\,d^2 y_i$.

In \eq{chi} we subtracted the term that corresponds to   the $ 2 \to 3$ transition at $ u(x_1,y_2) 
\,=\,1$. Indeed, at $u(x_1,y_2) =1$ this transition describes the decay of the colour dipole 
$(x_2,y_1)$ into two dipoles at any size of the dipole $(x_1,y_2)$ over which we integrate. Such 
a process has been taken 
into account in the first term of \eq{chi} which accounts for  $1 \to 2$ decays of all possible 
dipoles.
\footnote{In the first version of this paper as well as in the first version of Ref. \cite{LLTR} we 
made a mistake of  forgetting about this term. In doing so,  we incorrectly  generated the two 
Pomerons to 
one 
Pomeron 
transition. We thank all our colleagues  whose  criticism helped us to find a correct form for $2 
\to 3$ 
transition in \eq{ZEQ}.}

\eq{ZEQ} is a typical diffusion equation or Fokker-Planck equation \cite{GARD},   with the diffusion 
coefficient which depends on $u$. This is the master equation of our approach, and the goal of this 
paper is to find the asymptotic solution to this equation. In spite of the fact that this is a 
functional equation we intuitively feel; that this equation could be useful since we can develop a 
direct method for its  solution  and, on the other hand, there exist many  studies of such 
an equation 
in the literature ( see for example Ref. \cite{GARD}) as well as some physical realizations in 
statistical physics. The intimate relation between the Fokker-Planck equation, and the high energy 
asymptotic was first pointed out  by Weigert \cite{WE} in JIMWLK approach,  and has been discussed in 
Refs. 
\cite{BIW,IT1,MSW}.

 It should be stressed that in the case of leading $N_c$ approximation,  the master 
equation has only the first term with one functional derivative and, therefore, the Fokker-Planck 
equation degenerates to a Liouville's equation and describes the deterministic process,  rather than 
stochastic one which  the Fokker-Planck equation does. The solution to the  Liouville's equation is 
completely defined by the initial condition at $Y=0$,  and all correlations between dipoles are 
determined by the correlations at $Y=0$.  As has been shown \cite{LL,LLB,JANIK} that only by assuming 
that there are no correlations between dipoles at $Y=0$, we can replace the general Liouville equation 
by its simplified version, namely, by non-linear Balitsky-Kovchegov equation \cite{B,K}.

\begin{boldmath}
\subsection{$\Gamma\lb 2 \to 1 \rb$}
\end{boldmath}
For further use we need  more detailed information on the vertex, for merging  of two dipoles
in one dipole.

 As was shown in Refs. \cite{MSW,LLTR,IT2} the   vertex $\Gamma\lb 2 \to 1 \rb$ can be found from the 
integral 
equation which has the form
\beq \label{G21}
\int\,d^2\,x\,d^2\,y\, \Gamma_{ 2\rightarrow 1}\lb
x_1,y_1\,+\,x_2,y_2 \rightarrow x,y \rb
\,\gamma^{BA}\lb x,y;x',y'\rb\,\,=\,\,
\eeq
$$
=\,\,\,\int d^2\,x'_1,d^2\,y'_1\,,d^2\,x'_2,d^2\,y'_2\,\,\Gamma_{1 \rightarrow 2}\lb x',y'
\rightarrow
x'_1,y'_1 + x'_2,y'_2\rb\,
\,\gamma^{BA}\lb x_1,y_1; x'_1,y'_1\rb\,\,\gamma^{BA}\lb
x_2,y_2; x'_2,y'_2\rb
$$
where $\gamma^{BA}$ is a dipole-dipole elastic scattering amplitude in the Born approximation, which
is equal \cite{LI,BAA}
\beq \label{BA}
\gamma^{BA}\lb x,y; x',y'\rb\,\,=\,\,\frac{\bas^2}{2\,N^2_c}\,\frac{1}{16\,\pi^2}\,\ln^2 \lb \frac{(\x - 
\x')^2\,(\y -
\y')^2}{(\x - \y')^2\,(\y - \x')^2} \rb
\eeq

\eq{G21} is the basic equation from which the vertex $\Gamma_{2\rightarrow 1}$ can be extracted.
To achieve  this  we need to invert \eq{G21},  by acting on both sides of it by an operator which is  
inverse
to $\gamma_{BA}$ in operator sense. Fortunately, this operator is known to be a product of two
Laplacians \cite{LI,MSW,LLTR,IT2}:
\beq \label{LG21}
\Gamma_{2\rightarrow 1}\lb
x_1,y_1\,+\,x_2,y_2 \rightarrow x,y \rb\,\,=\,\,\frac{2 N^2_c}{\bas^2}\,\Delta_x\,\Delta_y\,
\,\,\int d^2\,x'_1,d^2\,y'_1\,,d^2\,x'_2,d^2\,y'_2\,\times
\eeq
$$
\,\Gamma_{1 \rightarrow 2}\lb x',y'
\rightarrow
x'_1,y'_1 + x'_2,y'_2\rb\,
\,\gamma^{BA}\lb x_1,y_1; x'_1,y'_1\rb\,\,\gamma^{BA}\lb
x_2,y_2; x'_2,y'_2\rb
$$
The exact evaluation of \eq{LG21}  is done in Ref. \cite{LLTR}, but for further presentation in this
paper we will need the vertex $\Gamma(2 \to 1)$ only in the form of \eq{LG21}.

\subsection{Scattering amplitude and its statistical interpretation}

As was shown in  Refs. \cite{K,LLB}) the scattering amplitude is defined as a functional
\begin{eqnarray}  \label{N}
N \lb Y;[\gamma_i] \rb  \,\,&=&\,\,- \,\,\int\,\sum^{\infty}_{n =1}
\,\frac{(-1)^n}{n!}\,\,\,\,\prod^n_{i=1}\,d^2 x_i\,d^2 y_i\,\gamma(x_i,y_i)\,\frac{\de}{\de 
u_i} Z\lb Y,[u_i] \rb|_{u_i =1} \nonumber\\
 \,\,&=&\,\,-\,\,\int \sum^{\infty}_{n =1}(-1)^n \,\,\prod^n_{i=1}\,d^2 x_i\,d^2 
y_i\gamma(x_i,y_i)\,\rho_n 
\lb r_1,b_1;\dots r_n,b_n;Y_0 \rb  
\end{eqnarray}
where $\gamma$ denotes an   arbitrary function which should be specified from the initial condition 
at 
$Y=0$. To calculate the amplitude we need to replace each term  $\prod^n_{i=1} \gamma(x_i,y_i)$  by
function  $\gamma_n (x_1,y_1; \dots; 
x_n,y_n)$  which  characterizes 
the amplitude   at low energies for simultaneous 
scattering of $n$ dipoles off the target.

From \eq{N} one can see that
\beq \label{RHO}
\rho_n\lb x_1,y_1; \dots ; x_n,y_n \rb \,\,\,=\,\,\frac{1}{n!}\,\prod^n_{i=1}\,\frac{\de}{\de 
\,\,u_i}\,Z\lb Y;[u_i] \rb |_{u_i=1}
\eeq

\begin{figure}[htb]
\begin{minipage}{10cm}
\begin{center}
 \includegraphics[width=0.80\textwidth]{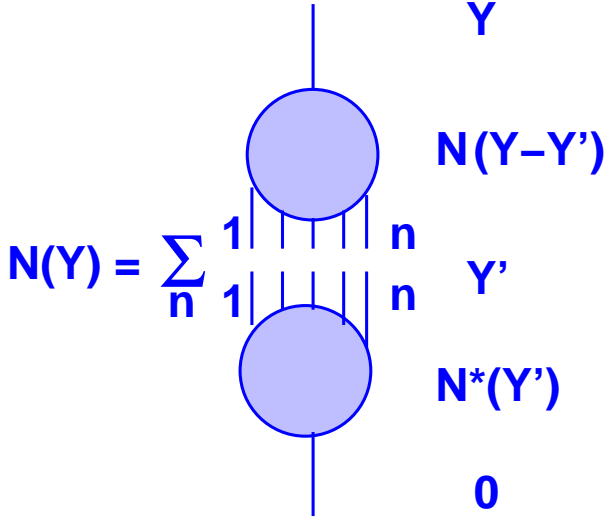}
\end{center}
\end{minipage}
\begin{minipage}{6cm}
\caption{\it The unitarity constraint in $t$-channel.}
\label{unita}
\end{minipage}
\end{figure}

\begin{figure}[htb]
\begin{minipage}{11cm}
\begin{center}
 \includegraphics[width=0.80\textwidth]{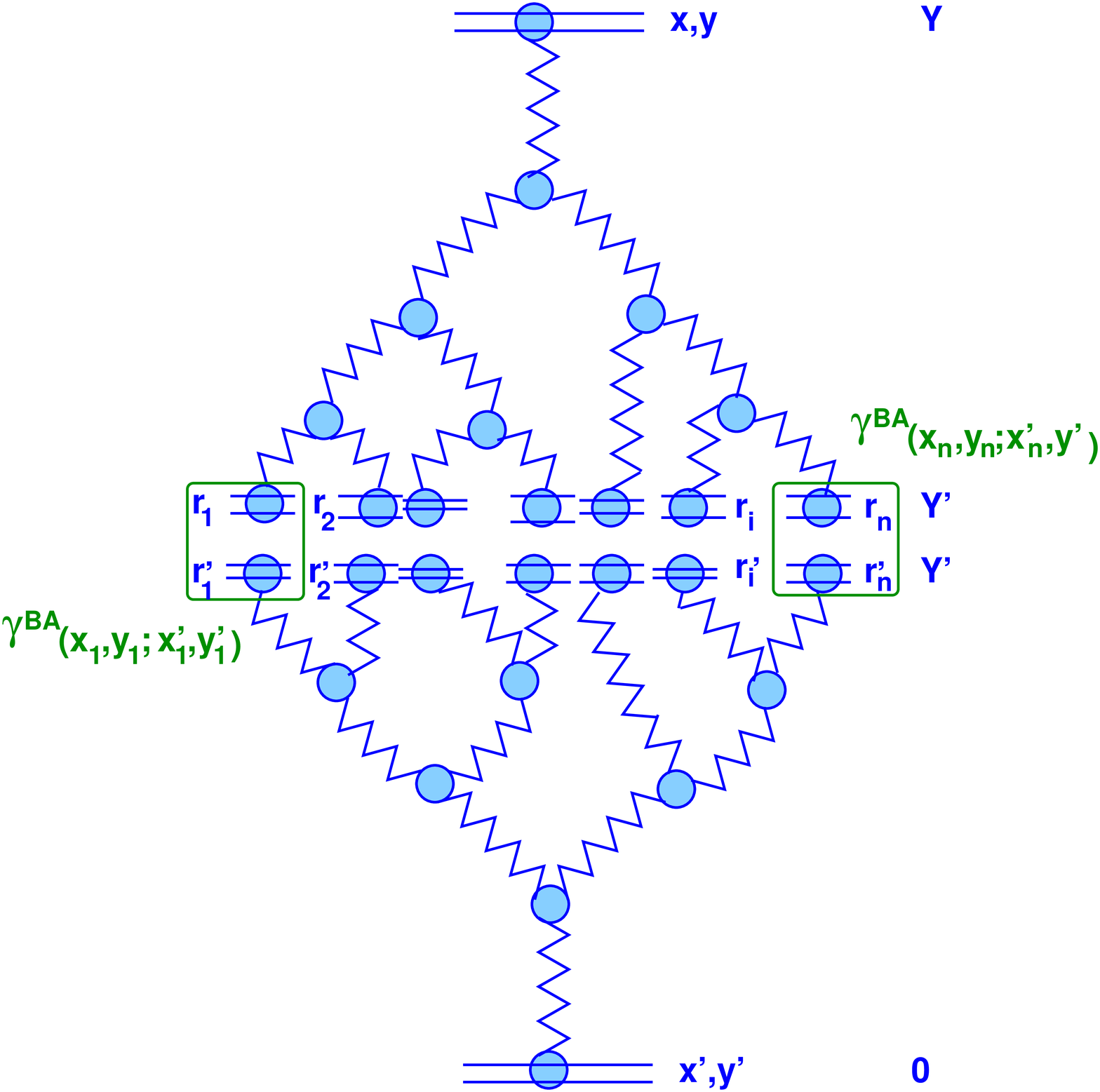}
\end{center}
\end{minipage}
 \begin{minipage}{5cm}
\caption{\it An example of enhanced diagrams that contribute to the unitarity constraint in 
$t$-channel.  Zigzag lines denote the BFKL Pomerons. This particular set of diagrams can be summed 
by the unitarity in $t$ channel and corresponds to Iancu-Mueller approach \cite{IM,KOLE}.}
 \label{gefunenh}
\end{minipage}
\end{figure}

As was shown by Iancu  and Mueller \cite{IM}  $t$-channel unitarity plays an  important if not 
crucial role in low $x$ physics (see also Refs. \cite{KOLE,MUSO}). In the context of this paper it 
 should be noted that  $t$-channel unitarity as a non-linear relation for the amplitude,  is 
able to determine the unknown parameters in the asymptotic solution.
  $t$ -channel unitarity for dipole-dipole scattering  can be written in the form (see 
\fig{unita})
\beq \label{TUNIT}
N(x,y; x',y';Y)\,\,=\,\,
\eeq
$$
\sum^{\infty}_{n=1}\,\,\int\,\prod^n_{i=1}\,d^2 x_i\,d^2 y_i\,
 \,\tilde{N}^p(x,y;\dots; x_i,y_i; \dots x_n,y_n; Y - Y')\,\tilde{N}^t(x',y';\dots;
x_i,y_i; \dots x_n,y_n; Y')\
$$
where $\tilde{N}^p$ and $\tilde{N}^t$ denote the amplitude of the projectile and target, 
respectively.
Actually, they are not exactly the amplitude. Indeed, accordingly \eq{TUNIT} their dimension should 
be 
$1/x^2$ while the amplitude $N$ is dimensionless.  On the other hand, we know that the unitarity 
constraint has a form $Im\,\,N\,\,=\,\,\sum_{n}\,N( 2 \to n)\,n^*(2 \to n)$ illustrated in \fig{unita}.
However, the amplitude in this relation should be taken in the  momentum representation, while we 
here consider 
the amplitude in the coordinate representation.
The difference is clear from \eq{N} where we have 
that the amplitude of interaction at low energy  $\gamma(x_i,y_i)$ has been taken into account 
in the definition of $N$. On the other hand in the unitarity constraints,  these amplitudes should 
only  be 
included  once  for  both amplitudes (see \fig{gefunenh}). The general way to do this 
is to re-write the unitarity constraints through the  function $\rho_n$ using \eq{RHO}.

It leads us to the following form of  the $t$-channel unitarity constraint:
$$ 
N\left(x,y,;x',y';Y\right)\,\,\,=
$$
\begin{eqnarray}
\,\,\,\,\,&=&\,\,\,\sum^{\infty}_{n=1}\,\,(-1)^n\,n!\,\int\,\,\prod^n_{i=1}\,\,
d^2\,x_i\,\,d^2\,y_i\,\,\,\prod^n_{i=1}\,\, d^2
x'_i\,\,d^2\,y'_i\,\,\,\gamma^{BA} (x_i,y_i;x'_i,y'_i)\,\, \label{TUNG}\\
 & &
\rho^p_n\left(x,y; x_1,y_1;\dots\,x_i,y_i;\dots ;\,x_n\,y_n,Y-Y' \right)\,\rho^t_n\left(x',y'; 
x'_1,y'_1;
\dots\,x'_i,y'_i;\,\dots;\,x'_n,y'_n; Y' \right)\,\,\nonumber
\end{eqnarray}
The factor $n!$ appears in \eq{TUNG}  is due to the fact that each dipole with rapidity $Y'$ from the 
target can interact   with any dipole from the projectile (see \fig{gefunenh}).

The unitarity constraint itself shows that the high energy amplitude could be described as a Markov 
process. Indeed, this constraint claims that the amplitude at later time ( $Y + \Delta Y $)  is 
determined entirely by the knowledge of the amplitude at the most recent time ($Y$) since $N(Y 
+\Delta  Y)= N(\Delta Y)\,N(Y)$ from  $t$-channel unitarity.

The relation between $Z$ and $N$ is very simple \cite{K}, namely
\beq \label{ZN}
N\lb  Y,[\gamma_i]\rb\,\,\,=\,\,1\,\,\,-\,\,\,Z\lb Y, [1 \,-\,\gamma_i] \rb
\eeq

Therefore, the amplitude $N$ is closely related to the characteristic function in the statistical 
approach. However, it is well known that it is better to use  the  cumulant generating function 
which is the logarithm of the characteristic function. In our case, we introduce the cumulant 
generating functional, namely,
\beq \label{PHI}
\Phi\lb Y;[\gamma_i] \rb \,\,\equiv\,\,\ln Z\lb Y;[\gamma_i] \rb \,\,=\,\,\sum^{\infty}_{n=1}\int 
\prod^n_{i=1}\,d^2 x_i \,d^2 y_i \gamma_i \lb x_i,y_i \rb\,p_n\lb\dots;x_i,y_i; \,\dots \rb  
\eeq
where $u_i = 1 - \gamma_i$. \eq{PHI} is the definition for functions $p_n$ which are related to 
cumulants. For example $p_2$ is related to $\langle|T(r_1)T_[r_2)| \rangle - \langle|T(r_1)\rangle 
\langle|T(r_2)\rangle$ rather than to $ \langle|T(r_1)T_[r_2)| \rangle$ as $P_n$ does. $T $ is the 
scattering operator.

The advantage of using $\Phi$ is the following: (i) if there are no correlations between dipoles  it  
is necessary and sufficient to keep only the first term in the series of \eq{PHI}; (ii) to take into 
account the two dipole correlations we need to keep the two first terms in \eq{PHI}; and (iii) the 
$n$-term 
in the series of \eq{PHI} describes the correlations between $n$-dipoles.

Substituting \eq{PHI} in \eq{ZEQ} we obtain the equation for $\Phi$
\beq \label{PHEQ}
\frac{\partial \Phi \lb Y;[\gamma_i] \rb}{\partial\,Y}\,\,=\,
\int d^4 q d^4 q_1 d^4 q_2\,\Gamma (q \to q_1 + q_2)\,\lb \gamma(q) 
\,-\,\gamma(q_1)\,\gamma(q_2)\,\rb\, \frac{\de}{\de \gamma(q)}\,\Phi\lb Y;[\gamma_i] \rb \,\,+
\eeq
$$
+\,\,\h \int d^4 q d^4 q_1 d^4 q_2\,\Gamma ( q_1 + q_2 \to q)\,\lb \gamma(q)
\,-\,\gamma(q_1)\,\gamma(q_2) \rb\,\times
$$
$$
\lb  \frac{\de}{\de \gamma(q_1)}\,\Phi\lb 
Y;[\gamma_i]\rb\,\frac{\de}{\de \gamma(q_2)}\,\Phi\lb Y;[\gamma_i] \rb
\,+\,\frac{\de^2}{\de \gamma(q_1)\,\de \gamma(q_2)}\,\Phi\lb Y;[\gamma_i] \rb \rb\,\,-
$$
$$
\,\,+\,\,\h \int \prod_{i=1}^2 d^4 q_i d^2 z\,\Gamma ( q_1 + q_2 \to 
(x_1,y_2) + (x_2,z)  + (z,y_1) )  
\, \gamma(x_1,y_2)\,\lb \gamma(x_2,y_1)\,\,-\,\,\gamma(x_2,z)\,\gamma(z,y_1)\,\,
\rb  \,\times
$$
$$
\lb \frac{\de}{\de \gamma(q_1)}\,\Phi\lb
Y;[\gamma_i]\rb \,\frac{\de}{\de \gamma(q_2)}\,\Phi\lb Y;[\gamma_i] \rb
\,+\,\frac{\de^2}{\de \gamma(q_1) \de  \gamma(q_2)}\,\Phi\lb Y;[\gamma_i] \rb
\rb
$$

where $u(q)\,\equiv\,1 \,-\,\gamma(q)$.

\section{Asymptotic solution in the toy-model}
\subsection{General description}
We start to solve the master equation (see \eq{ZEQ} and \eq{PHEQ})
by considering a simple toy model
 in which we assume  that interaction does not depend on the size of dipoles (see 
Refs. \cite{MUCD,LL,KOLE} for details).
For this model the master functional equation ( see \eq{ZEQ} ) degenerates into an ordinary 
equation 
in partial derivatives, namely
\beq \label{TM1}
\frac{\partial\,Z}{\partial Y}\,\,=
\eeq
$$
=\,\,- \Gamma(1 \rightarrow 2)\,\,u(1 - 
u) 
\,\frac{\partial\,Z}{\partial u}\,\,+\,\,\Gamma(2 \rightarrow 1)\,\,u^(1 - 
u)\,\,\frac{\partial^2\,Z}{(\partial u)^2}\,\,+\,\, \Gamma(2 \rightarrow 3)
u\,(1 - 
u)^2\frac{\partial^2\,Z}{(\partial u)^2}
$$

To obtain the scattering amplitude we need to replace $\gamma$ in \eq{ZN},  by the amplitude 
of 
interaction of the dipole with the target. 
For $N$ \eq{TM1} can be rewritten in the form:
\beq \label{TM2}
\frac{\partial\,N}{\partial Y}\,\,=
\eeq
$$
=\,
\,\, \Gamma(1 \rightarrow 2)\,\,\gamma(1 - 
\gamma)\frac{\partial\,N}{\partial \gamma}
\,\,+\,\,\Gamma(2 \rightarrow 1)\,\,\gamma (1 - \gamma)\,\,\frac{\partial^2\,N}{(\partial 
\gamma)^2}\,\,+\,\, \Gamma(2 \rightarrow 3)
\gamma^2\,(1 -\gamma)\frac{\partial^2\,N}{(\partial \gamma)^2}
$$
if $\gamma$ is small we can reduce \eq{TM2} to the simple equation
\beq \label{TM3}
\frac{\partial\,N}{\partial Y}\,\,=\,\,\Gamma(1 \rightarrow 2)\,\,\gamma 
\,\frac{\partial\,N}{\partial \gamma}
\eeq
which has the solution
\beq \label{TM4}
N\,\,=\,\,\gamma e^{ \Gamma(1 \rightarrow 2) Y}
\eeq
which describes the exchange of the Pomeron with the intercept $\Gamma(1 \rightarrow 2)$.

\begin{figure}[htbp]
\begin{center}
\epsfig{file= 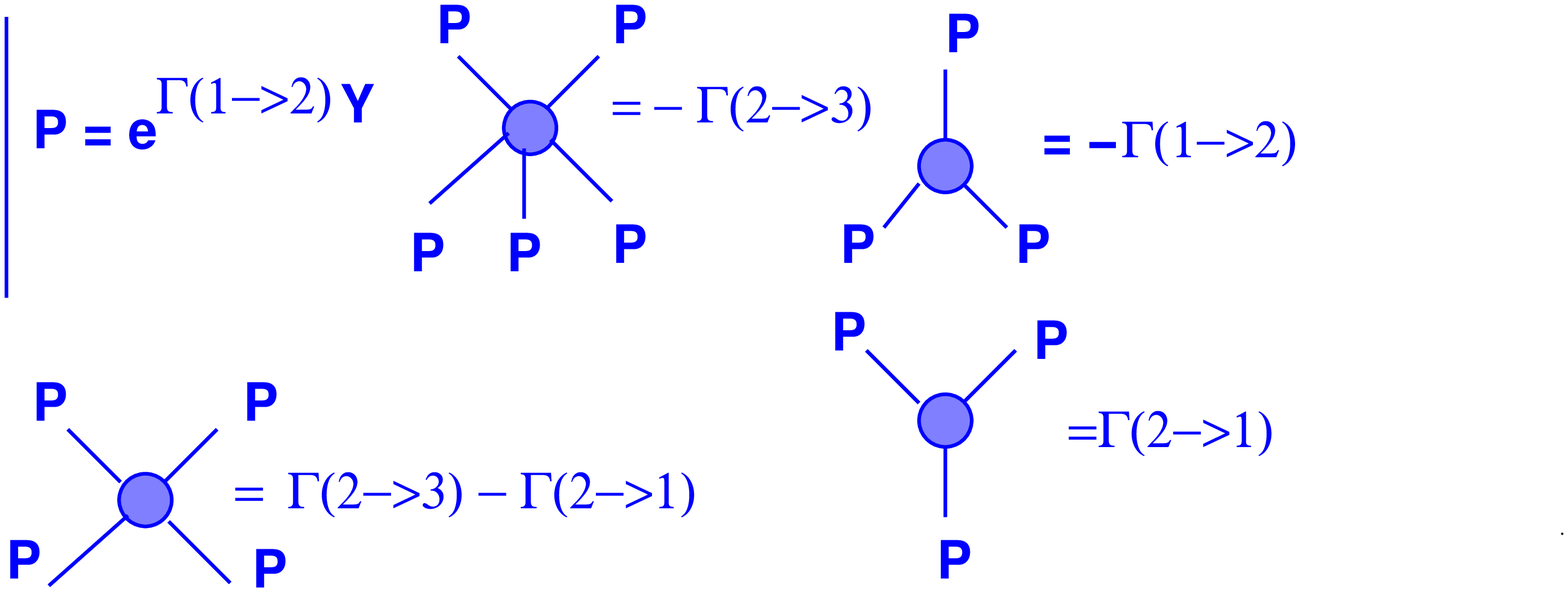,width=140mm}
\end{center}
\caption{\it Pomeron interactions described by \eq{TM2}.}
\label{pom}
\end{figure}

All other terms in \eq{TM2} have a very simple physical meaning, describing Pomeron 
interactions
(see \fig{pom}).

The probabilistic approach that we develop in this paper has an advantage that we can apply 
the  well developed formalism of the statistical physics  \cite{GARD}  to the partonic system (see 
Refs. 
\cite{BOPO,BIW,IT1,MSW} where statistical approach has been applied to our problems).
It is reasonable to do this if   we are dealing with a system with  a  large number of dipoles at 
high 
energy. We will try to illustrate this point below,  but first let us describe the strategy that we 
will follow in our search for a  solution.

\subsection{The strategy of our search for  the solution}
Generally speaking the solution to \eq{TM1} or to \eq{TM2}  can be written in the form
\beq \label{GENSOL}
Z(Y;u) \,\,=\,\,Z(\infty; u)\,+\,\Delta Z(Y,u)
\eeq
where asymptotic solutions  $Z(\infty; u)$ or $N(\infty; u)$,  are solutions of \eq{TM1} or \eq{TM2} 
with the l.h.s. equal 
to zero, namely,
it means that asymptotic form of $Z$ can be determined from the equation
\beq \label{NSA}
\,\, -\,\,\Gamma(1 \rightarrow 2)\,\,u\,(1 - u)\frac{\partial\,Z(\infty; u)}{\partial \,u}
\,\,+
\eeq
$$
+\,\,\Gamma(2 \rightarrow 1)\,\,u\, (1 -  u)\,\,\frac{\partial^2\,Z(\infty; u)}{\partial\,
u^2}\,\,+\,\, \Gamma(2 \rightarrow 3)
u\,(1 - u)^2 \frac{\partial^2\,Z(\infty; u)}{\partial\, u^2}= 0
$$
If we prove that the solution $\Delta Z(Y,u)$ which satisfies  the full \eq{TM1} decreases  at high 
energy ( at $Y \to \infty$) the asymptotic solution will give us the behaviour of the scattering 
ampllitude at high energy.

\eq{TM1} and \eq{TM2} are diffusion equations, in which the diffusion coefficient 
is a  function of $u$ or $\gamma$. Let us solve the simplified problem replacing \eq{TM1} by the 
equation with constant coefficient, namely,
\beq \label{GENSOL1}
\frac{\partial\,Z}{\partial {\cal  Y}}\,\,=
\,
\,\, -\,\frac{\partial\,Z}{\partial u}
\,\,+\,\, \frac{1}{\kappa}\frac{\partial^2\,Z}{\partial
u^2}
\eeq
where ${\cal Y} = \Gamma(1 \to 2)\,Y$ and $\kappa$ is defined in \eq{KAPPA}.

Then 
\beq \label{ASYMSOL}
Z(\infty,u) \,\,\,= \,\,\, e^{\kappa\,(u - 1)}
\eeq
 in which we fixed all constants,  assuming that 
all other solutions give small contributions at large values of $Y$ and  $Z(\infty;u=1)=1$.

The general solution for  \eq{GENSOL1} is known, namely, it is equal to
\beq \label{GENSOL2}
\Delta Z(Y;u)\,\,=\,\,\h\,\sqrt{\pi\,\kappa}\,\exp \lb \,-\,\kappa\, \frac{(u \,-\, {\cal 
Y})^2}{4\,{\cal 
Y}} \rb
\eeq
\eq{GENSOL2} is the Green function of \eq{GENSOL1},  and using it one can construct  the solution 
for  any 
initial condition. This solution vanishes at large ${\cal Y}$. This fact shows that 
\eq{ASYMSOL} is the asymptotic solution at high energy. 

Therefore, the strategy of searching for  the solution to our problem consists of two steps:
\begin{enumerate}
\item \quad Finding the asymptotic solution as a  solution to \eq{ZEQ} with zero l.h.s.;
\item \quad Searching for a solution to the full equation,  but assuming that this solution will be 
small at high energies ($Y\,\to\,\infty$).
\end{enumerate}
We also have to show that the solution, which we find,  satisfies  the initial condition which we
take  $Z(Y=0,u) = u$.

In the following  presentation we consider separately two cases $\Gamma(2 \to 3)=$ and $\Gamma(2 \to 
1) =0$
which lead us to further understanding of the asymptotic behaviour of the scattering  amplitude.

\begin{boldmath}
\subsection{Solution for $\Gamma(2 \to 3) =0$}
\end{boldmath}
\subsubsection{Asymptotic solution}

Considering a simple case with $\Gamma( 2 \to 3)=0$ we obtain a  solution for $Z(\infty;u)$ from 
\eq{NSA}
\beq \label{NGASA}
Z(u; Y \,\to\,\infty)\,\,=\,\,1 - B \,+\,Be^{\kappa (u - 1)}
\eeq
where
\beq \label{KAPPA}
\kappa \,=\,\,\frac{\Gamma(1 \,\to\,2)}{\Gamma(2
\,\to\,1)}\,\,=\,\,\frac{2\,N^2_c}{\bas^2}\,\,\gg\,\,1\,\,;\,\,\,\,\,\,\,\,\,
\tilde{\kappa}\,\,=\,\,\frac{\Gamma(1 \,\to\,2)}{\Gamma(2
\,\to\,3)}\,\,=\,\,2\,N^2_c\,\,\gg\,\,1
\eeq
It should be stressed that  in the toy model  we keep the same order of magnitude for the vertices 
as 
in the full QCD approach. This explains all numerical factors in \eq{KAPPA}. 
To avoid any
confusion we would like to draw
the reader  attention to the fact that we will use $\kappa$ and $\tk$ in further discussions to
denote
 the combination of
$\as$ and $N_c$ but not  the ratios of the vertices which are  functions  of coordinates in the
general case. These combinations give the order of magnitude for the ratios of vertices as far as the 
$\bas$ and $N_c$ factors  are  concerned.

\eq{NGASA} satisfies the initial condition that $Z(u=1;Y)\,=\,1$,  but the coefficient $B$ remains
undetermined, since we cannot use the initial condition at $Y=Y_0$. We will show below that the 
unitarity constraint in $t$-channel will determine $B=1$.
Expanding \eq{NGASA} we obtain:
\beq \label{NMOSA}
\lim_{Y \to \infty} P_0 (Y)\,\,=\, \,e^{- \kappa};\,\,\,\,\,\,\,\,\,\,\,\lim_{Y \to
\infty} P_n (Y)\,\,=\,\, \frac{\kappa^n}{n!}\,\,e^{- \kappa};
\eeq
We obtained the Poisson distribution with the average number of dipoles
$\langle n \rangle = \kappa\,\,\gg\,\,1$. Hence we will attempt  to discuss the evolution of our 
cascade using statistical methods \footnote{In
this subsection we follow Ref. \cite{BOPO} in which a wide class of  models for  interacting
Pomerons  is  considered.}.

To find the value of $B$ in \eq{NMOSA} we can use the unitarity constraint in $t$-channel,  which 
can be obtained from \eq{TUNG},  and which 
for 
the simple model reads as \cite{BOPO,IM,KOLE,LLB}
\beq \label{UNTM}
N(Y)\,\,=\,\,\sum^{\infty}_{n=1}\,(-1)^n\,n!\,\frac{1}{\kappa^n}\rho^p_n(Y - y)\rho^t_n(y)
\eeq
where $\rho$ both for the projectile and the target is defined by \eq{TUNIT}.
\eq{UNTM} corresponds to a simplification that in the toy model $\gamma^{BA}\,= 
\frac{1}{\kappa}\,\de( x_i - x'_i)\,\de( y_i - y'_i)$.

Rewriting \eq{UNTM} for $Y - y \,\gg\,1$ and $y \,\gg\,1$ and taking into account \eq{ZN}
we have
\beq \label{ININF}
N(\infty)\,=\,\sum^{\infty}_{n=1}\,(\frac{-1}{\kappa})^n \,n!\,\rho^p_n( \infty)\,\,\rho^t_n( 
\infty)
\eeq
Therefore,  we see that $B = 1$ since for large $\kappa$ \eq{ININF} reads as
$$
B \,=\,B^2 
$$

 Substituting $\rho^p_n( 
\infty)=\kappa^n_p/n!$ and $\rho^t_n(
\infty)=\kappa^n_t/n!$ we obtain
\beq \label{ININF1}
N(\infty)\,\,=\,\,1\,\,-\,\,e^{- \frac{\kappa^p\,\kappa^t}{\kappa}}\,\,=\,\,1\,\,-\,\,e^{- 
\kappa}
\eeq
where $\kappa^p$ and $ \kappa^t$ are related to the projectile and target, and they are equal 
$\kappa^p = \kappa^t\,=\,\kappa$ in our case. The unitarity constraint 
determines the value of $u$ in the solution of \eq{NGASA} which is relevant to the asymptotic 
behaviour of the amplitude. We can translate the result of  using  unitarity given by \eq{ININF1}, 
as the  $u = 0$ contribution  to the asymptotic behaviour of the amplitude, if we calculate it from 
\eq{NGASA}.

\subsubsection{Search for energy dependent solution}

\eq{NGASA}  leads us  to use the Poisson representation \cite{BOPO}:
\beq \label{PRE}
P(Y)\,=\,\int\,\,d \,\alpha\,F(\alpha,Y) \lb \frac{\alpha^n}{n!}\,e^{-\alpha} \rb \,\equiv \,\langle 
P_n(\alpha) \rangle
\eeq
where $< \dots >$ stands for averaging with weight $F(\alpha,Y)$.

The generating function $Z(Y;u)$ can be written as
\beq \label{PZSA}
Z(Y;u)\,\,=\,\,\int\,\,d \alpha\,\, F(\alpha,Y)\,e^{(u\,-\,1)\,\alpha}
\eeq
Our master equation (see \eq{TM1}) can be rewritten as
\beq \label{PFSA}
\frac{\partial\, F(\alpha,Y)}{\partial\,Y}\,\,=\,\,\,- \frac{\partial}{\partial
\,\alpha}\,\lb A(\alpha) \,F(\alpha,Y) \rb \,\,+\,\,\h \,\frac{\partial^2}{\partial
\,\alpha^2}\,\lb B(\alpha)\,F(\alpha,Y) \rb
\eeq
where
\begin{eqnarray}
A(\alpha) \,\,&=&\,\,\Gamma(1 \,\to\,2)\,\alpha \,-\,\Gamma(2 \,\to\,1)\,\,\alpha^2\,;
\label{FPA}\\
B(\alpha) \,\,&=&\,\,2\,A(\alpha);
\label{FPB}
\end{eqnarray}
\eq{PFSA} is an equation of Fokker-Planck type ( see Refs. \cite{BOPO,BIW,IT1,MSW}  with
positive diffusion coefficient $B(\alpha)$ at least for $\alpha < \kappa$. The
initial condition of \eq{ZIN1} gives the normalization of $F(\alpha,Y)$, namely,
\beq \label{FNORM}
\int\,d\,\alpha\,F(\alpha,Y)\,\,=\,\,1
\eeq

For positive $B(\alpha)$ the distribution function $F(\alpha,Y)$ is positive with the normalization
of \eq{FNORM}, and therefore can be considered as a probability distribution \cite{GARD}.
Hence, the energy dependence of the scattering amplitude can be given by averaging of
\eq{ININF1}, namely,
\beq \label{FINNSA}
N(Y)\,=\,
\eeq
$$=
\langle \langle N^{(\alpha, \bar{\alpha}} \rangle \rangle \,\equiv\,\int\,\,d
\alpha\,F(\alpha,Y-y)\,\,\int\,\,d \bar{\alpha}\,F(\bar{\alpha},y)\,\,\lb
1\,\,-\,\,e^{-\alpha\,\bar{\alpha}} \rb
$$

\eq{PFSA} is equivalent to the differential stochastic equation \cite{GARD}
\beq \label{STOCHEQ}
d \,\alpha\,=\,A(\alpha)\,+\,\sqrt{B(\alpha)}\,d\,W(Y)
\eeq
where $ d \,W(Y)$ is a stochastic differential for the Wiener process. Using \eq{STOCHEQ}, together
with the large number of dipoles involved in the process is the reason we hope 
 to solve the whole problem using the statistical approaches (see Refs. \cite{BOPO,IT1,MSW}).
However, it is too early to judge how successful this approach will be.

As one can see from \eq{STOCHEQ}  $\alpha$ grows until it reaches the zero of $B(\alpha)$ and it 
becomes frozen at this value at large $Y$. Therefore, we can solve \eq{PFSA} at large energy 
assuming that $\kappa - \alpha \,\ll\,\kappa$. In this limit $A(\alpha) $ can be replaced by 
$A(\alpha)\,=\,\kappa - \alpha = \xi$ and \eq{PFSA} reduces to the form
\beq \label{FKSI}
\frac{\partial F(\alpha)}{\partial \,{\cal Y}}\,\,= \,\, F(\alpha)\,\,(\xi\,+\,2)\frac{\partial 
F(\alpha)}{\partial 
\,\xi}\,\,+\,\,\xi\,\frac{\partial^2\,F(\alpha)}{\partial\,\xi^2}
\eeq
 where ${\cal Y}\, =\,\Gamma(1 \,\to\,2)\,Y$ with the solution
\beq \label{SOLFKSI1}
F(\alpha,{\cal Y})\,\,=\,\,\int\,\frac{d \,\omega}{2 \,\pi\,i}\,e^{\omega\,\,{\cal Y}}\,\,{\cal 
F}(\alpha, \omega)\,;
\eeq
and
\beq \label{SOLFKSI2}
{\cal F}(\alpha, \omega)\,\,=\,\,\phi(\omega) \,{}_1F_1 \lb 1 + \omega, 2, \xi \rb
\eeq
where ${}_1F_1$ is the confluent hypergeometric function of the first kind. We should choose the 
function $\phi(\omega)$ from the normalization condition of \eq{FNORM} and, finally,
\beq \label{FCALF}
{\cal  F}(\alpha, \omega)\,\,=\,\,\h {}_1F_1 \lb 1 + \omega, 2, \xi \rb
\eeq
To find the asymptotic behaviour of this solution at $Y \,\gg\,1$, reconsider 
\eq{SOLFKSI1} which has the form
\beq \label{SOLFF1}
F(\alpha,Y)\,\,=\,\,\int\,\frac{d \,\omega}{2 \,\pi\,i}\,e^{\omega\,\,\Gamma(1 \to 2)\, 
Y}\,\,{}_1F_1\lb 1 + \omega, 2, \xi \rb
\eeq
We can use  the integral representation for ${}_1F_1$ (see {\bf 9.211}(1) in Ref. \cite{RY}) 
and 
return to the Poisson representation of \eq{PZSA}.  Doing so,  the solution has the 
following form
\beq \label{SOLFF2}
Z(u,Y)\,=\,e^{\kappa (u - 1)}\,\Psi\lb{\cal Y} \,+\,\ln \lb \frac{1 + t}{1 - t} \rb \rb
\eeq
where $t \equiv \,2(u - 1)$ and the arbitrary   function $\Psi$ should be determined  from the 
initial 
condition.	One can see that from $Z(u=1;Y)=1$ this function is equal to 1.

This solution corresponds to the asymptotic solution which does not 
depend on $Y$ (see \eq{NGASA}). We have to consider a different region of $\alpha$ to find out 
how our system approaches the asymptotic regime given by \eq{NGASA}. Assuming that $\alpha 
\,\ll\,\kappa$ we can model $A$ and $B$ in \eq{PFSA} by $A(\alpha) = \alpha$. In this case
\eq{PFSA} has a simple form
\beq \label{SOLFF3}
\frac{\partial\, F(\alpha,Y)}{\partial\,{\cal Y}}\,\,=\,\,\,- \frac{\partial}{\partial
\,\alpha}\,\lb \alpha \,F(\alpha,Y) \rb \,\,+\,\, \,\frac{\partial^2}{\partial
\,\alpha^2}\,\lb \alpha\,F(\alpha,Y) \rb
\eeq
Returning to the variable $t = u -1$ (see \eq{PZSA}) we obtain the equation for $Z$ 
\beq \label{SOLFF4}
\frac{\partial\, Z(t,Y)}{\partial\,{\cal Y}}\,\,=\,\, - 2 \, Z(t,Y)\,- t(1 +t)\,\frac{\partial\, 
Z(t,Y)}{\partial\,t}
\eeq 
which has the solution
\beq \label{SOLFF5}
Z(t,Y)\,\,=\,\,\frac{(1 +t)^2}{t^2}\,\Xi\lb {\cal Y}\, + \,\ln \lb \frac{(1 +t)}{t}\rb \rb
\eeq
The solution of the master equation (see \eq{TM1}) can be written as
\beq \label{FSOL}
Z(t,Y)\,\,=\,\,e^{\kappa\,t}\,\,+\,\,\frac{(1 +t)^2}{t^2}\,\Xi \lb {\cal Y}\, + \,\ln \lb
\frac{(1 +t)}{t}\rb \rb
\eeq
Function $\Xi$ is arbitrary function and we determine  it from the initial condition that 
$Z(u,Y=0)=u$. 
Finally,
the solution has the form
\beq \label{FSOLF}
Z(t,Y)\,\,=\,\,e^{\kappa\,t}\, \,\,+\,\,e^{-2\, \Gamma(1 \to 2)\,Y}\, \lb \frac{1 +t}{1 + t 
-t\,e^{ 
- \Gamma(1 \to 2)\,Y}}\,\,-
\,\,\exp\{ \kappa\,\frac{t\,e^{- \Gamma(1 \to 2)\,Y}}{1 + t -t\,e^{
-
\Gamma(1 \to 2)\,Y}}\,\}\rb
\eeq

Therefore, the key difference between the case with only emission of dipoles (only $\Gamma(1 \to 2) 
\,\neq\,0$),  and the case when we take into account the annihilation of dipoles $\Gamma(1 
\to 2)\,\neq\,0$ and $\Gamma(2 \to 1)\,\neq\,0$, is  the fact that there exists an 
asymptotic solution which depends on $u$ (see \eq{NGASA}). 
\eq{FSOLF} shows that the energy dependent solution in the wide range of $t$ or $u$ satisfy the 
initial condition $Z(Y=0;u)=u$ and it decreases  at $Y \to \infty$. Therefore, \eq{ININF1} gives the 
asymptotic solution to our problem. 

\begin{boldmath}
\subsection{Solution for $\Gamma( 2 \to 1) = 0$}
\end{boldmath}
\subsubsection{Asymptotic solution}
For the case $\Gamma(2 \to 1)=0$ the system shows quite different behaviour, namely,
\beq \label{NGA1SA}
Z(u; Y \,\to\,\infty)\,\,=\,\,1 + \frac{B}{ 1 \,-\,\tilde{\kappa}}\,\lb (1 - u )^{1 - \tilde{\kappa}}
\,-\, (1 - u_0)^{1 - \tilde{\kappa}} \rb
\eeq
where $\tilde{\kappa}$ is given by \eq{KAPPA}.  Formally speaking we cannot satisfy the boundary 
condition $Z(u=1; Y \,\to\,\infty)$. The reason for this is obvious, since we cannot take  $u$ 
close to 1 and neglect $\Gamma \lb 2 \to 1 \rb $ term in \eq{TM1}. We can do this only for $u > u_o 
\approx 1 -  \tk/\kappa \propto \bas^2$. 

One can see that \eq{NGA1SA} gives
$$
 < n > = d Z(u; Y \,\to\,\infty)/d u |_{u =1} = B (1 - u_0)^{- \tk} B ( \tk/\kappa)^{- \tk} 
\,\,\gg\,1
$$
Therefore, once again  we  hope that the statistical approach can 
 work in this system.

\eq{NGA1SA} is the asymptotic solution in this case. Once more we should use the unitarity 
constraint to fixed the parameter $B$. Repeating all calculations that led to \eq{ININF1} we obtain
\beq \label{ININF2}
N(\infty)\,=\,\,\sum^{\infty}_{n =1}\,\frac{1}{n!}\,\frac{\Gamma^2( \tk + n)}{\Gamma^2( 
\tk)}\,\lb\,-\,\frac{1}{\kappa\,(1 - 
u_0)^2} \rb^n\,\, \equiv
\eeq
$$
 \equiv \,\, {}_2F_0\lb \tk,\tk; ; -\,\frac{1}{\kappa\,(1 -
u_0)^2} \rb \,\,<
$$
$$
<\,1 - \exp\lb - \frac{1}{\kappa\,(1 -
u_0)^2} \rb\,\,+\,\,\exp\lb\,-\,\frac{1}{\kappa\,(1 -
u_0)^2}\rb \,\,Ei \lb - \kappa (1 - u_o)^2 \rb
$$
 where $\Gamma(x)$ is the Euler gamma function, ${}_2F_0$ is the generalized hypergeometric 
function\cite{RY}     and $ Ei(x) = \int_{- \infty}^{x} 
\frac{e^{t}}{t} dt $ is exponential integral.  Considering $1 - u_0 \approx\,-  \tk/\kappa$ we
see that 
\beq \label{ININF3}
N(\infty)\, \,\,<\,\,1 - \exp\lb - \frac{\kappa}{\tk^2}\rb\,\,+\,\exp\lb  \frac{\kappa}{\tk^2} \rb
\,Ei \lb - \frac{\kappa}{\tk^2} \rb
\eeq
\begin{boldmath}
\subsubsection{Solution for $\Delta Z(Y; u)$}
\end{boldmath}

 As
has been shown in \eq{NGA1SA}, the asymptotic  distribution at large values of Y is not the 
Poissonian
type. If we try to introduce the distribution function $F(\alpha,Y)$, the equation for it has the 
form
\beq \label{F23}
\frac{\partial\, F(\alpha,Y)}{\partial\,\tilde{Y}}\,\,=
\eeq
$$
\,\,- \frac{\partial}{\partial
\,\alpha}\,\lb \tilde{A}(\alpha) \,F(\alpha,Y) \rb \,\,+\,\,\h \,\frac{\partial^2}{\partial
\,\alpha^2}\,\lb \tilde{B}(\alpha)\,F(\alpha,Y) \rb\,\,-\,\,\frac{\partial^3}{\partial\,\alpha^3}
\lb \tilde{C}(\alpha)\,F(\alpha,Y) \rb
$$
where $\tilde{Y} \,=\,\Gamma(1\,\to\,2)\,Y$ and
\begin{eqnarray}
\tilde{A}(\alpha) \,\,&=&\,\,\alpha \,-\, \,\frac{1}{\kappa}\,\,\alpha^2\,;
\label{FPTA}\\
\tilde{B}(\alpha) \,\,&=&\,\,2\,\alpha \,-2\,\,\lb
\,\,\frac{1}{\kappa}\,\,-\,\,\frac{1}{\tilde{\kappa}}\,\,\rb\,\alpha^2\,;
\label{FPTB} \\
\tilde{C}(\alpha) \,\,&=&\,\,\frac{1}{\tilde{\kappa}}\,\,\alpha^2\,; \label{FPTC}
\end{eqnarray}
$\tilde{B}(\alpha)$ does not have a zero and, therefore, we expect that the asymptotic behaviour 
will be related to the large values of $\alpha$. For the  master equation (see \eq{TM1}) it means 
that  $u\,\to\,1$ will be essential. In the simple case $\Gamma(2 \to 1)=0$ and $u \,\to\,1 (1 - u 
\,\ll\,1$ the 
master equation degenerates to
\beq \label{TF1}
\frac{\partial Z(u,Y)}{\Gamma(1 \to 2) \partial Y}\,=
\eeq
$$
\,-\,u( 1 - u)\,\frac{\partial 
Z(u,Y)}{\partial\,u}\,\,+\,\,\frac{( 1 - u)^2}{\tilde{\kappa}}\,\frac{\partial^2 
Z(u,Y)}{\partial\,u^2}\,\,\equiv\,(1 - \frac{1}{\tilde{\kappa}}) \frac{\partial 
Z(\zeta,Y)}{\partial\,\zeta}\,\, 
+\,\frac{1}{\tilde{\kappa}}\,\frac{\partial^2 Z(\zeta,Y)}{\partial\,\zeta^2}
$$
where $\zeta = \ln (1 - u)$. The solution of this equation is simple, namely,
\beq \label{TF2}
Z(\zeta,Y)\,\,=\,\,\int^{a + i \infty}_{a - i \infty}\,\,\frac{d \lambda}{2\,\pi\,i}\,{\cal 
Z}(\lambda)\,e^{ \lb (1 - 
\frac{1}{\tilde{\kappa}})\,\lambda\,+\,\frac{1}{\tilde{\kappa}}\,\lambda^2 
\rb \Gamma(1 \to 2)\,Y\,+\,\lambda\,\zeta}
\eeq
where ${\cal Z}(\lambda)\,=\,1/(\lambda - 1)$ using  the initial condition that $Z(u,Y=0)=u$.

The solution which satisfies our  initial condition is equal to
\beq  \label{TF3}
Z(\zeta,Y)\,\,=\,\,erf\lb \frac{\eta}{2\,\sqrt{ \Gamma(1\to 2)Y/\tk}} , \rb
\eeq
where  $ \eta \,=\,(1 + \frac{1}{\tilde{\kappa}})\,\Gamma(1\to 2)\,Y\,\,+\,\,\zeta$ and 
$erf(x)$ is the error function given by $erf(x) = ( 2/\sqrt{\pi})\,\int^x_0\,\,\exp( - t^2 )\,d t$.

At $Y =0$ \eq{TF3} leads to $Z(\zeta,Y=0) = 1 \,\approx\, u$,   since we assume that $u$ is close to 
unity.

At $Y \to \infty$ we have 
\beq \label{TF4}
Z(\eta, Y\,\to\,\infty )\,\,\to\,\,1\,\,-\,\,\sqrt{\frac{4 \,\Gamma( 1 \to 2)\,\,Y}{\pi\,\tk\, \,\eta 
} }\,\,\,\exp \lb - \frac{\eta^2\,\tk}{4 \,\Gamma( 1 \to 2)\,\,Y} \rb
\eeq

Therefore, we have obtained a  solution and the only problem, that remains, is the behaviour of 
the 
asymptotic solution at $u \,\to\,1$. 
 We need to investigate the region of $u \to 1$, but we will do this using 
a new approximation, in which we use the large parameters of our approach ( see \eq{KAPPA}).

\begin{boldmath}
\subsection{Semi-classical approach for large $\kappa$ and $\tilde{\kappa}$}
\end{boldmath} 
Considering  the solution of \eq{FSOLF},  we  notice that the function $\Phi$ (see \eq{PHI} and 
\eq{PHEQ}) for this solution is large and it is proportional to $\kappa$. This observation triggers 
a search for a  semi-classical solution assuming that $ \Phi^2_{\gamma} >> \Phi_{\gamma,\gamma}$
where $\Phi_{\gamma} \equiv\,d Phi/d \gamma$. 

For $\Gamma(2 \to 3)=0$ in the toy model \eq{PHEQ} has the form
\beq \label{PHEQTM}
\frac{\partial\,\Phi(Y,\gamma)}{\partial \,{\cal Y}}\,=\,\,\gamma ( 1 - \gamma)\,\{ 
\frac{\partial\,\Phi(Y,\gamma)}{\partial 
\,\gamma}\,\,+\,\,\frac{1}{\kappa}\lb \frac{\partial\,\Phi(Y,\gamma)}{\partial 
\,\gamma} \rb^2 \}
\eeq
Fist, we see that the asymptotic solution $\Phi(\infty;\gamma) = - \kappa\,\gamma$ is the same as 
that we 
obtained 
in
\eq{NGASA} and \eq{FSOLF}.  We  try to find the solution to \eq{PHEQTM} for all values of $Y$
in the form $\Phi({\cal Y};\gamma)\,\,=\,\,\Phi(\infty;\gamma)\,\,+\,\,\Delta\Phi({\cal Y};\gamma)$ 
considering $\Delta \Phi \,\,\ll\,\,\Phi(\infty;\gamma)$.
For such a solution \eq{PHEQTM} reduces to a linear equation
\beq \label{LPHEQ}
\frac{\partial\,\Delta\Phi(Y,\gamma)}{\partial \,{\cal Y}}\,=\,\,-\,\gamma ( 1 
- \gamma)\,\frac{\partial\,\Delta\Phi(Y,\gamma)}{\partial
\,\gamma}
\eeq
The solution to this equation has the form
\beq \label{SC1}
\Delta \Phi(Y,\gamma) \,\,=\,\,H \lb {\cal Y} \,+\,\,\ln \lb \frac{\gamma -1}{ 
\gamma} \rb \rb
\eeq
where $H$ is an arbitrary function which should be found from the initial conditions $Z(Y=0,[u_i]) = 
u$.

Assuming that $\Delta \Phi \,\ll\,\Phi(\infty,\gamma)$ at $Y=0$,  we can reconstruct the solution 
given by \eq{SC1}, namely,
\beq \label{SC2}
Z(\gamma,Y)\,\,=\,\,\exp \lb \Phi(\infty,\gamma)\,+\,\Delta \Phi(Y,\gamma) \rb\,\,=\,\,
\eeq
$$
\frac{1 - \gamma}{1 -  \gamma +  \gamma\,e^{- \Gamma(1 \to 2)Y}}\,\exp\{ - \kappa\,\gamma \,\,+\,\, 
\kappa\,\frac{\gamma e^{- \Gamma(1 \to 2)Y}}{1 -  \gamma +  \gamma\,e^{- \Gamma(1 \to 2) Y}} \}
$$
This solution differs from the solution given by \eq{FSOLF}, but leads to the same behaviour at 
large values of $Y$. The difference  is  obvious  since $\Delta \Phi$ is not small at 
small 
values of $Y$.

We   attempt to find a solution using the same methods as for the case  $\Gamma(2 \to 1)=0$.

The asymptotic solution for $\Phi$ is $\Phi(\infty; u )\,\,=\,\,-\tk\,ln(\gamma)$. This can be seen 
directly from \eq{NGA1SA}, and  it also appears as the solution to the following equation
\beq \label{23PHI}
0\,\,=\,\,\gamma\, ( 1 - \gamma)\,\{
\frac{\partial\,\Phi(\infty; \gamma)}{\partial
\,\gamma}\,\,+\,\,\frac{1}{\tk}\,\gamma\,\lb \frac{\partial\,\Phi(\infty; 
\gamma)}{\partial
\,\gamma} \rb^2 \}
\eeq
For $\Delta \Phi(Y,u)$ ($ \Phi(Y,u)\,\,=\,\,\Phi(\infty; u)\,\,+\,\,\Delta \Phi(Y,u)$)
we have
\beq \label{23PHI1}
\frac{\partial\,\Delta\Phi(Y; \gamma)}{\partial \,{\cal Y}}\,=\,\,\,\gamma\, ( 1
- \gamma)\,\frac{\partial\,\Delta\Phi(Y; \gamma)}{\partial
\,\gamma}
\eeq

A general solution to this equation is 
\beq \label{23PHI2}
\Delta \Phi(Y,\gamma)\,\,=\,\,S\lb {\cal Y} \,- \,\ln \lb \frac{\gamma}{\gamma - 1} \rb\,\rb
\eeq
where $S$ is an arbitrary function which should be determined from the initial conditions $Z(Y=0;u) = 
u$ 
which translates into initial conditions for $\Delta \Phi$ as
\beq \label{23PHI3}
\Delta \Phi(Y=0,u)\,\,=\,\,\tk \,\ln (\gamma)\,\,+\,\ln(1 - \gamma)
\eeq

Finally, we obtain the solution in the form:
\beq \label{23PHI4} 
Z({\cal Y};\gamma)\,\,=\,\,1 \,\,+\,\frac{ 1 - \gamma}{ 1 - \gamma + \gamma 
e^{-{\cal Y}}} \,\,\lb\,1\,\,-\,\,\gamma \,\,-\,\,\gamma\,e^{-{\cal Y}} \rb^{ - \tk}\,\,
\eeq

This is the  solution which tends to the asymptotic solution as $Y \to \infty$, at least at small 
values 
of $\gamma \,<\, 1/\tk$. Note, we have not achieved the correct normalization $Z(u =1,Y)=1$.
 However, we believe this is connected to the  problem of ill defined limit of $\Gamma(2 \to 1)$, for 
small $\gamma \ll 
\tk/\kappa$. To get a better understanding  the situation better we consider the general case in the 
semi-classical 
approach.

\subsection{Semiclassical approach to a general case}

It is easy to obtain the asymptotic solution
\beq \label{SCG1}
\Phi( \infty, \gamma)\,\,=\,\,-\,\tk\,\ln \lb 1\,\,+\,\,\frac{\kappa}{\tk}\,\gamma \rb 
\eeq
 $\Delta \Phi(Y,u)$ ($ \Phi(Y,u)\,\,=\,\,\Phi(\infty; u)\,\,+\,\,\Delta \Phi(Y,u)$) has the form
\beq \label{SCG2}
\frac{\partial\,\Delta\Phi(Y; \gamma)}{\partial \,{\cal Y}}\,=\,\,\,\gamma\, ( 1
- \gamma)\,\frac{\partial\,\Delta\Phi(Y; \gamma)}{\partial
\,\gamma}
\eeq
and the solution to this equation has the same form as \eq{23PHI2}, but with a different initial 
condition:
\beq \label{SCG3}
\Delta\Phi(Y=0; \gamma)\,\,=\,\,\tk \,\ln  \lb 1\,\,+\,\,\frac{\kappa}{\tk}\,\gamma \rb
\,\,+\,\,\ln \lb 1 \,-\,\gamma \rb
\eeq

Using  the  arbitrary function $S$ ( see  \eq{23PHI2} ) and  from \eq{SCG3} we, finally,  
obtain the 
answer for the generating functional $Z$
\beq \label{SCGZ}
Z\lb Y, \gamma \rb\,\,=\,\,\frac{ (1 - \gamma) \,e^{\Gamma(1 \to 2)\,\,Y}\,}{ \gamma \,+\,( 1 - 
\gamma)\,e^{\Gamma(1 \to 2)\,\,Y}}\,\,\lb 1\,\,+\,\,\frac{\kappa}{\tk}\,\gamma \rb^{- \tk}
\,\,\lb 1 \,+\,\frac{\kappa}{\tk}\,\frac{\gamma}{\gamma \,+\,( 1 - \gamma)\,e^{\Gamma(1 \to 2)\,\,Y}} 
\rb^{\tk}
\eeq

This solution satisfies all requiments: (i) $Z\lb Y,u =1 \rb \equiv Z\lb Y,\gamma =0 
\rb \,=\,1$; (ii) $Z\lb Y=0,u \rb = u \equiv 1 - \gamma$; and at $Y \to \infty$ $
Z\lb Y,u \rb\,\,\to\,\,Z(\infty,u)$.

The asymptotic solution of \eq{SCG1} leads to the following $Z$
\beq \label{SCGZA}
Z\lb \infty, \gamma \rb\,\,=\,\,\lb\,1\,\,+\,\,\frac{\kappa}{\tk}\,\rb^{ - \tk}
\eeq
Using \eq{RHO} we obtain
\beq \label{SCG4}
\rho_n\,\,=\,\,\frac{(-1)^n}{n!}\,\frac{\Gamma(\tk + n)}{\Gamma(\tk)}\,\,\lb \frac{\kappa}{\tk} \rb^n
\eeq
This $\rho_n$ is the same as for the case of $\Gamma ( 2 \to 1)=0$ (see \eq{NGA1SA}) 
if $ 1 - u_0$ is chosen to be equal to $1 - u_0\,\,=\,\,\tk/\kappa$. 
 It should be stressed that \eq{SCGZ} and \eq{SCG4}  lead to the scattering 
amplitude given  by \eq{ININF2}. It shows that we  correctly guessed  the value of $1 - 
u_0$. Only for 
$ 1 - u
\,\leq 1 - u_o$  
 do we have to 
take into account the process of merging of two Pomerons into one Pomeron. In other words, we can 
neglect the $ 2 \to 1$ process if we put the initial  condition for the generation 
functional $ Z(Y,u)$ at $u = u_0$, namely $Z(Y,u = 1 - \tk/\kappa) =1$. This property resembles and 
even supports the idea of Ref. \cite{IMM},  that merging processes suppress  correlations when these 
correlations are small.

\subsection{Lessons for searching a general solution}

 We view this toy model as a training ground to help us find   a general solution, 
and as an aid  suggesting  directions for such a search. We learned 
several lessons that will be useful:
\begin{enumerate}
\item \quad Our approach in  searching for  the solution consists of the following steps:
\begin{itemize}
\item \quad Finding the asymptotic solution as the solution to the master equation ( see 
\eq{ZEQ}, \eq{PHEQ}, \eq{TM1} and \eq{TM3}) with zero l.h.s.;
\item \quad Using the large parameters of our theory given by                                                  
\eq{KAPPA},  we can develop the 
semi-classical approach for searching both for the asymptotic solution and for the correction to this 
solution,  
that provide the form  of the generating functional approaching  its asymptotic value;
\item \quad The corrections to the asymptotic solution decrease at large values of $Y$, and can be 
found 
from  the Liouville-type linear equation;
\item \quad The important region  of $u$ ($\gamma$) are $u \,\to 1$ and $\gamma \,\to\,0$, which 
should 
be specified by using the unitarity constraint;
\end{itemize}
 \item \quad The inclusion of $\Gamma(2 \to 3)$ is very crucial  for the form 
of 
solution; 
\item \quad The asymptotic form of the solution  differs from the solution to the 
Balitsky-Kovchegov equation: amplitude $N$ does not approach 1 ($N \rightarrow 1$) at $Y \rightarrow  
\infty$, but rather $N(\infty,\gamma)$ is a function of $\gamma$. In particular, it means that we do 
not expect  geometrical scaling \cite{GS}  for the solution to the full set of equations. We also do 
not 
expect  that the scattering amplitude to show a black disk behaviour on  reaching the value of unity 
at high energies. We predict the   gray disc behaviour for this scattering amplitude;
 \item \quad It is easy to find the solution using the equation for the cumulant generation 
functional $\Phi$ (see \eq{PHI} and \eq{PHEQ}).
\item \quad The role of the $2 \to 1$ process is very interesting: this process suppresses the small 
values of $\gamma \,\leq \tk/\kappa$ in a such way that they can be neglected,  supporting the idea 
of 
Ref. \cite{IMM};
\end{enumerate}

It should be stressed  that we found such a  solution to the simplified evolution equation (see \eq{TM1}), which has   
the form of $Z(\infty,u) \,+\,\Delta Z(Y,u)$ with $\Delta Z(Y,u) \,\to\,0$ at large $Y$. It is shown that $\Delta Z(Y,u)$ 
satisfies the initial condition for the generating functional. Therefore, for the toy model we prove that our solution is 
the solution to the evolution equation if we believe that we have the only one solution. For the solution to the general 
evolution equation (see \eq{ZEQ} ) we will not be able to show that our solution satisfies the initial condition for  the 
generating functional. Therefore, it is possible that could exist a different solution which we overlooked in our 
approach but which will satisfy the initial condition while ours does not. This is the reason, why the solution to the 
evolution equation in the toy model, in which we can check that such situation does not occur, is so important for our 
approach.  

It should be stressed that all new feathures of the solution appear only in the next to leading order in $1/N_c$. In the 
leading order the JIMWLK-B approach leads to a solution with behaves as a black disc. 

At first sight,  the gray disc behaviour of the scattering amplitude  looks unrealistic. However, the situation is just 
opposite, the gray disc behaviour is rather natural in the parton model  (see Ref. \cite{Kancheli}) and  we have spent a  
decade to 
understand 
how it is possible that we have a black disc behaviour for the scattering 
amplituide in QCD \cite{FR}. The  black disc behaviour  stems from the simple fact that in JIMWLK-B approach the 
number of `wee" dipoles of each size increases as power of energy due to BFKL emission. It means that the dipoles started 
to interact with each other even if this interaction  is small (proportional to $\as$). At very high energy the 
probability to find any dipole is equal to unity. The Pomeron loops lead to diminishing of the BFKL Pomeron intercept 
and, therefore, to  a suppression of the number of the `wee' partons. Finally, the increase of the number of `wee' 
dipoles stops before the probability to find a dipole reaches 1.  The entire picture seems to be close to so called 
critical Pomeron scenario (see Ref. \cite{CRPOM}): the only theoretical model of Pomeron that has been solved.

Armed with this knowledge  we will now  to solve the general equations ( see \eq{ZEQ} and /or 
\eq{PHEQ} ).

\section{Asymptotic solution  (general consideration)}

\begin{boldmath}
\subsection{Solution for $\Gamma(2 \to 3)$=0}
\subsubsection{Solution at $Y \,\to\,\infty$}
\end{boldmath}

Our strategy in searching for the solution will be based on the lessons we learnt  from the toy 
model. First we
try to solve the master equation (see \eq{PHEQ} ) assuming that
\begin{eqnarray}
\,\frac{\de^2 \Phi}{\de \gamma(x_1,y_1) \de  \gamma(x_2,y_2)} &\ll &
  \,\frac{\de \Phi}{\de \gamma(x_1,y_1)}\frac{\de \Phi}{ \de
\gamma(x_2,y_2)} \label{AS1} \\
\Phi\lb \infty;[\gamma_i] \rb &=&- \kappa\,\, \int\,d^2 x\,d^2 y\,\,\Delta_x\,\Delta_y
\phi(x,y;\infty)\,\,\,\gamma(x,y)
\label{AS2}
\end{eqnarray}
Substituting \eq{AS2} into \eq{PHEQ} and  putting    the l.h.s. of \eq{PHEQ} to zero,   we obtain the
following equation for asymptotic solution $\phi(x,y;Y=\infty)$:
\beq \label{AS3}
0\,\,=\,\,\int\,d^2\,z\,\,K(\x,\y;\z)\,\Delta_x\,\Delta_y\,\phi(x,y;\infty)\,\,-
\,\,\Delta_x\,\Delta_y \lb
\,\,\int\,d^2 z\,\,K(\x,\y;\z)\,\,\int\,d^2 x_1\,d^2\,y_1\,\,\right.
\eeq
$$
\left.
\gamma^{BA} \lb \x , \z; \x_1\,\y_1 \rb
\Delta_{x_1}\,\Delta_{y_1} \phi(x_1,y_1;\infty)\,\,
\int\,d^2 x_2\,d^2y_2\,\,
\gamma^{BA} \lb
\x , \z;
\x_2\,\y_2 \rb
\Delta_{x_2}\,\Delta_{y_2} \phi(x_1,y_1;\infty)\rb
$$

where $K$ is given by \eq{K} and we used \eq{LG21} for the vertex $\Gamma(2 \to 1)$.
Integrating \eq{AS3} by parts with respect to $x_1,y_1$ and $x_2,y_2$ and using
\beq \label{AS4}
\Delta_{x_1}\,\Delta_{y_1} \gamma^{BA} \lb \x , \y; \x_1\,\y_1 \rb\,\,\,=\,\,\de \lb \x - \x_1 \rb
\de \lb \y - \y_1 \rb
\eeq
we find that function $\phi(x,y;\infty)$ satisfies the equation
\beq \label{ASPH}
0\,\,=\,\,\int\,d^2z\,K(\x,\y;\z)\,\lb ( \phi(x,y;\infty)\,\,-\,\,\phi(x,z;\infty)\,\phi(z,y;\infty)
\rb
\eeq
To obtain \eq{ASPH} from \eq{AS3} we need to
 assume that
\beq \label{ASUM}
\int\,d^2\,z\,\,K(\x,\y;\z)\,\Delta_x\,\Delta_y\,\phi(x,y;\infty)\,\,\approx\,
\,\Delta_x\,\Delta_y\,\int\,d^2\,z\,\,K(\x,\y;\z)\,\,\phi(x,y;\infty)
\eeq
Indeed, as we show below $\Delta_x\,\Delta_y\,\phi(x,y;\infty)$  is very singular and behaves as $
\Delta_x\,\Delta_y\,\phi(x,y;\infty) \propto\,\de \lb \x - \x_1 \rb
\de \lb \y - \y_1 \rb$. The function $\int\,d^2\,z\,\,K(\x,\y;\z)$ is less singular. Therefore, as
far as the most singular part of solution is concerned \eq{ASUM} holds.

The detailed  discussion of the solution to \eq{AS3} will appear in  a separate paper \cite{KLBK} and
here we
will only  use   the fact that $\phi(x,y;\infty) =1$ satisfies  this equation.

Therefore, the cumulant generating functional for the asymptotic solution is equal to
\beq \label{ASPHI}
\Phi\lb Y;[\gamma_i] \rb \,\,=\,\,\kappa\,\, \int\,d^2 x\,d^2 y\,\,\Delta_x\,\Delta_y
\phi(x,y;Y)\,\,\,\gamma(x,y)
\eeq
Using \eq{ASPHI} and unitarity constraint of \eq{TUNG},  we can find the asymptotic behaviour for
the
scattering amplitude. One can see  that the answer is
\beq \label{ASN}
N\lb x,y;x',y';\infty \rb \,\,=\,\,1\,\, -
\eeq
$$
-
\,\,\exp \lb  - \kappa^2\,\int\,\,d^2\,\bar{x} \,d^2\,\bar{y}
d^2\,x"\,d^2\,y"
\Delta_{\bar{x}}\,\Delta_{\bar{y}}\,\phi(x,y;\bar{x},\bar{y},\infty)\,\gamma^{BA}(\bar{x},\bar{y};x",y")
\Delta_{x"}\,\Delta_{y"}\phi(x',y';x",y",\infty) \rb
$$
Using the explicit form for $\gamma^{BA}$ as well as \eq{AS4},  we can rewrite \eq{ASN} in the form:
\beq \label{FASN}
N\lb x,y;x',y';\infty \rb \,\,=\,\,1\,\, -
\eeq
$$
\,\,\exp \lb -
\kappa^2\frac{\bas^2}{2\,N^2_c}\,\int\,\,d^2\,x" \,d^2\,y"
\,\phi(x,y;x",y";\infty))\,\Delta_{x"}\,
\Delta_{y"}\phi(x',y';x",y";\infty) \rb
$$
where $\phi$ is the solution of  \eq{ASPH}.
 
\subsubsection{Approaching the asymptotic solution}
As  in the toy model we will search for a solution in the following form:
\beq \label{APA1}
\Phi\lb Y;[\gamma_i] \rb \,\,= \,\,\Phi\lb \infty;[\gamma_i] \rb + \Delta \Phi\lb Y;[\gamma_i]
\rb\,\,
\eeq
assuming \eq{AS1},  and neglecting the contribution $\propto \,(\Delta \Phi)^2$.

The resulting equation for $\Delta \Phi\lb Y;[\gamma_i]\rb$ has the form
\beq \label{APA2}
\frac{\partial\,\,\Delta \Phi\lb Y;[\gamma_i]\rb}{\partial\,{\cal Y}}
\,\,=\,\,
\,\,\int\,d^2 x,d^2 y d^2\,z\,d^2 x_1\, d^2 y_1\,d^2 x_2\, d^2 y_2 \,\,
\eeq
$$
\lb u(x,y)
- u(x_1,y_1)\,u(x_2,y_2)\rb
\lb\,\,K(\x,\y;\z)\,\de \lb \x - \x_1 - \x_2 \rb \,\de \lb \y - \y_1 - \y_2 \rb\,\,-\,\,
2\,\,\Delta_{x}\,\Delta_{y}\,\,K(\x,\y;\z) \,\,\times \right.
$$
$$
\left.
\gamma^{BA} \lb \x , \z; \x_1\,\y_1 \rb\,\,
\Delta_{x_1}\,\Delta_{y_1} \phi(x_1,y_1;\infty)\,\,
\gamma^{BA} \lb
\x , \z;
\x_2\,\y_2 \rb \rb
\frac{\partial\,\,\Delta \Phi\lb Y;[\gamma_i]\rb}{\partial\,\gamma(x_2,y_2)}
$$
\eq{APA2} is a Liouville-type equation which can  easily be  solved assuming that the functional
$\Delta
\Phi\lb Y;[\gamma_i]\rb \,=\,\Delta \Phi\lb [\gamma_i ({\cal Y};x_i,y_i)]\rb$ . Using this
assumption we
can re-write
$$
\frac{\partial\,\,\Delta \Phi\lb {\cal Y};[\gamma_i]\rb}{\partial\,{\cal
Y}}\,\,=
\,\,\int\,d^2x\,d^2\,y
\frac{\partial\,\,\Delta \Phi\lb {\cal Y};[\gamma_i]\rb}{\partial\,\gamma({\cal
Y};x,y)}\,\,\frac{\partial\,\gamma \lb  {\cal  Y};x,y\rb}{\partial\,{\cal Y}}
$$
and
reduce \eq{APA2} to the following form:
\beq \label{ASA3}
\frac{\partial\,\gamma\lb  {\cal  Y};x,y\rb}{\partial\,{\cal Y}}\,\,=
\eeq
$$
\,\,
 \int\,\,d^2 \,z\, K(\x,\y;\z)\,\lb u({\cal Y};\x,\y) \,-\,u({\cal Y};\x,\z)\,u({\cal
Y};\y,\z)\,\rb
 \,\,-\,2\,\kappa\,\Delta_x\,\Delta_y\,\int\,\,d^2\,z \,d^2\, x_2 \,d^2\,y_2
K(\x,\y;\z)\,\,\times
$$
$$
\gamma^{BA}(\x,\z; \x_2,\y_2)\,\phi(x,z;\infty) \lb \,u({\cal Y};x,y))
\,-\,\,u({\cal Y}; x_2,y_2)\,\,u({\cal Y}; x,z) \rb
$$
where function $u = 1 - \gamma$ and  $\gamma\lb  {\cal  Y};x,y\rb$ are  determined by the initial
condition at $Y=0$,
namely, at $Y= 0$ we have only one dipole with coordinate $x,y$ with
\beq \label{ASAIN}
\Delta \Phi\lb {\cal Y} = 0;[\gamma_i]\rb\,\,=\ln(1 - \gamma({\cal Y}=0,x,y)) \,+\,\kappa \,
\gamma({\cal Y}=0,x,y)
\,\Delta_x\,\Delta_y\,\phi(x,y,\infty) \,\approx
\eeq
$$
\,\,\approx\,\,-\,\,\gamma({\cal Y}=0,x,y)\,+\,\kappa
\gamma({\cal Y}=0,x,y)
\,\Delta_x\,\Delta_y\,\phi(x,y,\infty)\,\,\equiv\,\, \Pi(x,y)\,\,\gamma({\cal Y}=0,x,y)
$$

Using \eq{ASAIN} we can re-write \eq{ASA3} in terms of $\Delta \Phi\lb {\cal Y} = 0;[\gamma_i]\rb$.
Neglecting terms of  order  $(\Delta \Phi)^2$  we obtain the following equation

\beq \label{ASA5}
\frac{\partial\,\Delta \Phi \lb  {\cal  Y};x,y\rb}{\partial\,{\cal Y}}\,\,=
\eeq
$$
\,\,
 \int\,\,d^2 \,z\, \lb\,\,K(\x,\y;\z)\,
-\
\,2\,\kappa\,\Delta_x\,\Delta_y\,\int\, \,d^2\, x_2 \,d^2\,y_2
K(\x,\y;\z)\,\,
\gamma^{BA}(\x,\z; \x_2,\y_2)\,\phi(y,z;\infty)\,\,\rb \times
$$
$$
 \lb\, 2\,\Pi(x,y)\frac{\Delta
\Phi({\cal
Y};x,z)}{\Pi(x,z)}\,\,-\,\, \Phi({\cal Y};x,y)\,\rb
$$
\eq{ASA5} can be solved   in log approximation. The first term in this equation is 
the
familiar BFKL linear equation, therefore, we  need to evaluate  the second term in \eq{ASA5}.

First,  we assume that at high  energies the nonlinear corrections set the new scale : the
saturation momentum $Q_s(Y)$ \cite{GLR,MUQI,MV}  and the integration which we have in the equation
is really cut off at
$|\x - \z|$ and /or $|\y - \z|$ smaller than $1/Q_s$ \cite{LT,MU02}. Actually,  $\phi(y,z;\infty)$
reaches a  maximum
value $\phi(y,z;\infty) = 1$ which is a solution to \eq{ASPH}.
We can rewrite $K(x,y;z)$ in the form
$$
K(x,y;z)\,\,\equiv\,\,\Delta_z\,\ln \lb \frac{( \x - \z)^2}{(\y - \z)^2} \rb
$$
Integrating by parts with respect to  $z$, and differentiating $\gamma^{BA}$ with respect to $x$ 
only,
 we find that  the second term has the form
\beq \label{2TERM}
2\,\kappa\,\Delta_x\,\Delta_y\,\int\, \,d^2\, x_2 \,d^2\,y_2
K(\x,\y;\z)\,\,
\gamma^{BA}(\x,\z; \x_2,\y_2)\,\phi(y,z;\infty)\,\,
 2\,\Pi(x,y)\frac{\Delta
\Phi({\cal
Y};x,z)}{\Pi(x,z)}\,\,=
$$
$$
\,\,2\,\kappa\Delta_y\,\int\, \,d^2\, x_2 \,d^2\,y_2
\ln \lb \frac{( \x - \z)^2}{(\y - \z)^2} \rb\,\Delta_x\,\Delta_z
\gamma^{BA}(\x,\z; \x_2,\y_2)\,\,\, 2\,\,\Pi(x,y)\frac{\Delta
\Phi({\cal
Y};x_2,y_2)}{\Pi(x_2,y_2)}\,=
\eeq
$$
=\,\,2\,\int\,\,d^2z\,\frac{1}{(\y - \z)^2}\,\Delta\Phi({\cal
Y};x,z)
$$
where we put $\Pi(x,y)  = \Pi(x,z)$ to  within the logarithmic accuracy.

Collecting both terms and using \eq{AS4},  we obtain the following equation for $\Delta \Phi$
\beq \label{ASAFIN}
\frac{\partial\,\Delta \Phi \lb  {\cal  Y};x,y\rb}{\partial\,{\cal Y}}\,\,
=\,\,- \,\int\,\,d^2\,z\,\frac{1}{(\y - \z)^2} \lb \,2\, \Delta \Phi \lb  {\cal  Y};x,z\rb
\,\,-\,\,\Delta \Phi \lb  {\cal  Y};x,y\rb \,\rb
\eeq
This is  the BFKL equation. However,  we solve this equation assuming that $\ln[ (\x -
\y)^2\,Q^2_s)]\,\,\gg\,\,1$. Considering $|\x - \z| \,\ll\,|\x - \y|$  we reduce \eq{ASAFIN} to the
following equation
\beq \label{ASAFIN1}
\frac{\partial\,\Delta \tilde{\Phi} \lb  {\cal  Y};x,y\rb}{\partial\,{\cal Y}}\,\,=\,\,\int^{(\x -
\y)^2}_{\rho^2}\,\,\,\frac{d^2 \,(x - z)}{(\x - \z)^2}\,\,\lb\,2\,\Delta \tilde{\Phi} \lb  {\cal
Y}x,y\rb \rb
 \eeq
where $ \Delta \tilde{\Phi}\lb  {\cal  Y};x,y\rb =  \Delta \Phi \lb  {\cal  Y};x,y\rb /(\x - \y)^2$.

Actually, the typical cutoff $\rho$ is of the order of $\rho^2 = 1/Q^2_s(Y)$ in the saturation
region \cite{LT,MU02}.
\eq{ASAFIN1} can be re-written in the differential form
\beq \label{ASAFIN2}
\frac{ \partial^2\,\Delta \tilde{\Phi} \lb  {\cal
Y};x,y\rb}{\partial\,Y\,\,\partial\,\ln((\x-\y)^2\,Q^2_s(Y))} \,\,=
\eeq
$$
=\,\,- \Delta \tilde{\Phi}\lb
{\cal Y};x,y\rb \,\,-\,\,\ln((\x-\y)^2\,Q^2_s(Y))\,\,\frac{ \partial\,\Delta \tilde{\Phi}\lb
{\cal Y};x,y\rb}{\partial\,\ln((\x-\y)^2\,Q^2_s(Y))}
$$

To find the solution to \eq{ASAFIN2}
 we change the variable introducing a new variable
\cite{LT,MUTR,BOKOLE,KOLE}
\beq \label{NVZ}
z\,\,=\,\,\ln((\x - \y)^2\,Q^2_s(Y))\,\,=\,\,\bas\,\frac{\chi(\gamma_{cr})}{1 -
\gamma_{cr}}\,Y\,+\,\ln((\x - \y)^2\,Q^2_0)
\eeq
where $\gamma_{cr}$ is determined by the following equation \cite{GLR,MUTR,BOKOLE}
\beq \label{GAMCR}
\frac{\chi(\gamma_{cr})}{1 -
\gamma_{cr}}\,\,=\,\,-\,\,\frac{ d\,\,\chi(\gamma_{cr})}{d\,\,\gamma_{cr}}
\eeq
In terms of the new variable,  \eq{ASAFIN2} has the form
\beq \label{ASAFIN3}
\frac{d^2 \Delta \tilde{\Phi}(z)}{d\,z^2}\,\,=\,\,- \frac{1 -
\gamma_{cr}}{\chi(\gamma_{cr})}\,\lb \Delta \tilde{\Phi}(z)\,\,-\,\,z\,\frac{d \Delta
\tilde{\Phi}(z)}{d\,z} \rb
\eeq
and
the solution of \eq{ASAFIN3} is
\beq \label{APSOL1}
\Delta \Phi \,\,=\,\,\Pi(x,y)\,\gamma(x,y)\,\,\exp\lb - \frac{1 - \gamma_{cr}}{2\,\,\chi(
\gamma_{cr}))}\,\,z^2
\,\rb
\eeq

To our surprise the asymptotic behaviour of $\Delta \Phi$  turns out to be the same, as 
 the asymptotic behaviour of  the Balitsky-Kovchegov equation \cite{LT}.

Using the unitarity constraint of \eq{TUNG},  and repeating all the calculations that led us to
\eq{FASN}
we obtain,  that the main corrections to the asymptotic behaviour given by \eq{FASN}  have the form:
$$
N\lb x,y;x',y',Y\rb \,\,=\,\,N\lb x,y;x',y',\infty\rb\,\,
\lb \,1\,-\,\,\int\,d^2 x" \,d^2
y"\,\Pi(x,y;x",y") \Pi(x',y';x",y")\,\times \right.
$$
\beq \label{ASPFIN}
\left.
\lb \exp \lb  -\, \bas\, \frac{1 - \gamma_{cr}}{2\,\,\chi(
\gamma_{cr}))}\,\,(Y -Y')^2 \rb \,\,+\,\,\exp \lb  - \, \bas\,\frac{1 - \gamma_{cr}}{2\,\,\chi(
\gamma_{cr}))}\,\,(Y')^2 \rb \rb \rb
\eeq
We assume in \eq{ASPFIN} that  $Y'$ and $Y -Y'$ are much larger than $ \ln ( x - y)^2\,Q^2_0$ or $
\ln
( x' - y')^2\,Q^2_0$ . For large $Y$ the minimal corrections occur at $Y' = \h Y$ . Therefore
\beq \label{ASPFIN1}
R_{\kappa}\,\,=\,\,\frac{N_{\kappa}\lb x,y;x',y',Y\rb\,\,-\,\,N_{\kappa}\lb
x,y;x',y',\infty\rb}{N_{\kappa}\lb
x,y;x',y',\infty\rb}\,\,\propto\,\,\exp \lb\, - \, \bas\,\frac{1 - \gamma_{cr}}{8\,\,\chi(
\gamma_{cr}))}\,\,(Y)^2\, \rb
\eeq
where we denote  by $\kappa$ the fact that we consider the example with   $\Gamma(2 \to 3)=0$.

It is interesting to compare this result with the solutions that have been found :
\begin{itemize}
\item \quad \,For the Balitsky - Kovchegov equation the ratio  of \eq{ASPFIN1} is of the order of
\beq \label{R1}
R\,\,=\,\,\propto\,\,\exp \lb \, - \, \bas\,\frac{1 - \gamma_{cr}}{2\,\,\chi(
\gamma_{cr}))}\,\,(Y)^2\,\rb\,\,;
\eeq
\item \quad \, For Iancu-Mueller \cite{IM}  approach which can be justified only in the limited
range of
energies we have
\beq \label{R2}
R\,\,=\,\,\propto\,\,\exp \lb \, - \, \bas\,\frac{1 - \gamma_{cr}}{4\,\,\chi(
\gamma_{cr}))}\,\,(Y)^2\,\rb\,\,;
\eeq
\end{itemize}
Comparing \eq{R1} and \eq{R2} with \eq{ASPFIN1} we see that the process of merging of two dipoles
into one dipole  crucially  change  the approach to the asymptotic behaviour .

\subsubsection{Main results}

The main results of our approach are given by \eq{ASPFIN} and \eq{ASPFIN1}. These equations show 
that at high energies the scattering amplitude approaches the asymptotic solution which is a 
function of coordinates ($N\lb x,y;x',y',\infty\rb$, see \eq{FASN}). This function is smaller than 1 
($ N\lb x,y;x',y',\infty\rb\,<\,1$) as it should be due to the unitarity constraints \cite{FROI}.

The corrections to the asymptotic behaviour turn out to be small (see \eq{ASPFIN1}). They 
are suppressed 
as $\exp ( - C Y^2)$, but the coefficient $C$ is found to be  4 times less than for the solution of 
the 
Balitsky-Kovchegov equation.

The toy model has  proved to be very useful,  and we  showed that this model can serve as a 
guide,  since we reproduce  all the main features of the toy model solution, in the general solution 
to 
the problem. It should be stressed that the semi-classical approach based on a  large parameter 
$\kappa 
\,\gg\,1$, gives the correct approximation to the problem.   

\begin{boldmath}
\subsection{General consideration}
\subsubsection{ Solution at $Y \,\rightarrow\,\infty$ for $\Gamma(2 \to 1) = 0$}
\end{boldmath}
As in section 4.1.1  we  search for  the asymptotic solution, as the solution 
to the master equation (see \eq{PHEQ}) with zero l.h.s.
Assuming \eq{AS1} and using \eq{KAPPA} for $\tilde{\kappa}$ 
we try to solve \eq{PHEQ} looking for a solution of the form
\beq \label{A231}
\Phi(\infty,[u_i])\,\,=\,\,- \,\tk \,\ln \lb 
\frac{\int\,\,d^2\,x\,d^2\,y\,\,\Theta(x,y)\,\,\gamma(x,y)}{\int\,\,d^2\,x\,d^2\,y\,\,\Theta(x,y)} 
\rb\,
\eeq
$$
=\,\,\,\,- \,\tk \,\ln \lb \frac{{\cal F}}{N} \rb
$$
where $\Theta$ is a function which is equal to 1 for $|\x| < R$ and $|\y| < R$ and  $\Theta = 0$ 
for $|\x| > R$ and $|\y| > R$ where $R$ is the 
largest scale in the problem, and 
\beq \label{A232}
{{\cal F}}(\infty,[\gamma_i])\,\,=\,\,\int\,\,d^2\,x\,d^2\,y\,\,\Theta(x,y)\,\,\gamma(x,y)
\,\,\,\mbox{and} \,\,\,N\,\,=\,\,\int\,\,d^2\,x\,d^2\,y\,\,\Theta(x,y)
\eeq

Substituting in \eq{PHEQ} we obtain the following equation 
$$
0\,\,=\,\,-\,\tk\,\int \,d^2\,x\,d^2\,y\,d^2\,z\, \Gamma( (x,y) 
\to (x,z) + (z,y))\,\lb \,\gamma(x,y) \,-\,\gamma(x,z)\,\gamma(z,y) \rb\,\frac{ \Theta(x,y)}{{\cal 
F}}\,\,- 
$$
$$
\,\,\tk^2\,\int \,d^2\,x_1\,d^2\,y_1\,d^2\,x_2\,d^2\,y_2\,\,d^2\,z \,\,
\Gamma( (x_1,y_1) + (x_2,y_2) \to (x_1,z) + (z,y_2) + (x_2,y_1) )\,\,\times
$$
\beq \label{A233} 
\,\lb \,\gamma(x_1,y_2)\,\gamma(x_2,y_1)\, \,-\,\gamma(x_1,z)\,\gamma(z,y_2)\,\gamma(x_2,y_1) 
\rb 
\frac{ 
\Theta(x_1,y_1)}{{\cal F}}\,\frac{
\Theta(x_2,y_2)}{{\mathcal F}} 
\eeq

Since $\Gamma( (x_1,y_1) + (x_2,y_2) \to (x_1,z) + (z,y_2) + (x_2,y_1) )\,=\,\tk\,\,\Gamma( 
(x_1,y_2) \to (x_1,z) + (z,y_2))$,  these two terms cancel each other if we denote 
the variable of integration in the first term as $ (x_1,y_1)$  or as $ (x_1,y_2)$
instead of $(x,y)$.

From \eq{A231} we obtain the generating functional at $Y \to \infty$, which is equal
\beq \label{Z23}
Z(Y,[\gamma_i]) \,\,=\,\,\lb \,\frac{\int\,d^2\,x\,d^2\,y\,\Theta(x,y)\,\gamma(x,y)}{N} 
\,\rb^{- \tk}
\eeq
One can see that \eq{Z23} cannot  be normalized by the condition $ Z(Y,[u_i=1]) =  Z(Y,[\gamma_i=0]) 
=1$. Indeed, as for the case of the toy model (see \eq{NGA1SA}) $Z(Y,[\gamma_i])$ diverges at small 
$\gamma$'s  
and the
value of $\gamma_{0,i}$ should be fixed by including $\Gamma( 2 \to 1) \neq 0$.
Including $\Gamma( 2 \to 1)$ we obtain 
 the same normalization condition,  but at
$$ \gamma_i\, = \,\gamma_{0,i} \approx\,\,\tk/\kappa 
\,\ll\, 1 \,.$$ 
The solution that satisfies this modified initial condition 
($Z(Y,[\gamma_i=\gamma_{0,i}]) \,=\,1$) has the  form
\beq \label{ZF23}
Z(Y,[\gamma_i]) \,\,=\,\,\lb \,\frac{\int\,d^2\,x\,d^2\,y\,\Theta(x,y)\,\gamma(x,y)}{{\cal F}( \infty; 
[\gamma_{0i}]) }
\,\rb^{- \tk}
\eeq

Using \eq{RHO} and \eq{ZF23}  one calculates $ \rho_n$ which are
\beq \label{A23RHO}
\rho_n(x_1,y_1; \dots\; x_b,y_n;Y=\infty)\,\,=
\,\,\prod^n_{i=0}\,\,(-1)^n\,\frac{1}{n!}\,\,\frac{\Gamma( - 
\tk)}{\Gamma(- \tk - 
n)}\, 
\,\frac{\Theta(x_i,y_i)}{{\cal F}(\infty;[\gamma_{0,i}])}
\eeq

 We will find $\gamma_{0,i}$  in the 
next subsection.

Using \eq{RHO} and the unitarity constraint of \eq{TUNG} we can find the asymptotic behaviour for 
the scattering amplitude. 
\beq \label{A23SOL}
N\lb(x,y;x',y'; \infty \rb \,\,\,=
\eeq
$$
=\,\,{}_2F_0 \lb \tk,\tk;\, ; - \frac{\,\int\,d^2 
\,\hat{x}\,d^2\,\hat{y}\,d^2 
\tilde{x}\,d^2\,\tilde{y}\,\,\Theta(\hat{x},\hat{y})\,\,\Theta(\tilde{x},\tilde{y})\,\,\gamma^{BA} 
\lb \hat{x},\hat{y}; \tilde{x},\tilde{y} \rb}{{\cal F}^2(\infty;[ \gamma_{0,i}])} \rb
$$
where ${}_2F_0$ is the generalized hypergeometric function \cite{RY}.

To obtain the answer we need to estimate the value of $\gamma_{0}$.

\begin{boldmath}
 \subsubsection{ $\gamma_0(x,y)$}
 \end{boldmath} 
As we have learned from the toy
model, we need to compare the second term of \eq{PHEQ} or \eq{ZEQ}, with the third term in
these equations using the asymptotic solution of \eq{A231}.  $\gamma_0(x,y)$ can be determined from
the equation which equates the second term to the third one.  Namely, we have 
\beq \label{GA01}
\int\,d^2 x\,d^2 \,y \,\gamma_0(x,y)\,\,\Gamma_{2 \to 1}\lb (x, z) + (z,y)  \to (x,y)\rb d^2 z
\,\,\,=\,\,\int\,d^2 x_1\,d^2 \,y_1 \,d^2 x_2\,d^2 \,y_2
\eeq
$$
\,\gamma_0(x_1,y_2)\,\,\gamma_0(x_2,y_1) \,\,\Gamma_{2 \to 3}\lb (x_1,y_1) + (x_2,y_2)  \to (z,y_1) +
(x_2,z) + (x_1, y_2) \rb d^2 z $$ 
Using the explicit forms of the vertices (see \eq{V23} and
            \eq{LG21} ) we can re-write \eq{GA01} in a more convenient form, namely 
\beq \label{GA02}
\frac{1}{\kappa}\int\,d^2 x\,d^2 \,y \gamma_0(x,y)\,\,\Delta_x\,\Delta_y \int
d^2\,z\,\,K(\vec{x},\vec{y};\vec{z})\,\,\gamma^{BA}(x,z; x_1 y_1) \,\gamma^{BA}(z,y; x_2 y_2)\,\,=
\eeq
 $$
 =\,\,\frac{1}{2\,\tk}\,\,\gamma_0(x_1,y_2)\,\,\gamma_0(x_2,y_1)\,\,\int\,\,\lb
K(\vec{x_2},\vec{y_1};\vec{z})\, \,+\,\, K(\vec{x_1},\vec{y_2};\vec{z}) \rb d^2\,z $$
 Taking
$\Delta_{x_1}\,\Delta_{y_1}\,\Delta_{x_2}\,\Delta_{y_2}$ from both sides of \eq{GA02} and integrating 
 over $x_1,x_2,y_1,y_2$ we can reduce
this equation to a simpler form using the following properties of the Born amplitude
 \beq \label{GA03}
\Delta_{x}\Delta_{y}\,\gamma^{BA}(x,z; x_1 y_1)\,\,=\,\,\,\delta^{(2)}( \vec{x} - \vec{x}_1 )  
\,\delta^{(2)}( \vec{y} - \vec{y}_1 )\,\,+\,\,\delta^{(2)}( \vec{x} - \vec{y}_1 )\,\delta^{(2)}(
\vec{y} - \vec{x}_1 )
 \eeq 
Indeed, we reduce \eq{GA02} to the form 
\beq \label{GA04}
\frac{1}{\kappa}\,\int\,d^2\,x\,d^2\,y \,\, \Delta_{x}\Delta_{y}\, 
\gamma_0(x,y)\,\,\int\,d^2\,z\,K\lb \vec{x},  \vec{y}; \vec{z} \rb\,\,\,=
\eeq
$$
=\,\,
\frac{1}{\tk}\,\int\,d^2\,x_1\,d^2\,y_1\,d^2\,x_2\,d^2\,y_2
\Delta_{x_1}\Delta_{y_2} 
\gamma_0(x_1,y_2)\,\,\Delta_{x_2}\,\Delta_{y_1} \,\lb \,\gamma_0(x_2,y_1)\,\,\int\,d^2\,z\,K\lb 
\vec{x}_2,  
\vec{y}_1; 
\vec{z} \rb\,\rb
$$

Approximating
\beq \label{APR}
\Delta_{x_2}\,\Delta_{y_1} \,\lb \,\gamma_0(x_2,y_1)\,\,\int\,d^2\,z\,K\lb
\vec{x}_2,
\vec{y}_1;
\vec{z} \rb\,\rb\,\,\approx\,\, \Delta_{x_2}\,\Delta_{y_1} \, \,\gamma_0(x_2,y_1) \times
\lb 
\int\,d^2\,z\,K\lb \vec{x}_2,
\vec{y}_1;
\vec{z} \rb \rb
\eeq
we obtain that
\beq \label{APR1}
\int\,d^2\,x\,d^2\,y \,\, \Delta_{x}\Delta_{y}\,\gamma_0(x,y)\,\,=\,\,\frac{\kappa}{\tk}
\eeq

The solution to \eq{APR1} is
\beq \label{GA05}
\gamma_0(x,y)\,\,\,=\,\,\frac{1}{16\,\pi^2}\,\frac{\tk}{\kappa}\,\,\ln^2 \frac{x^2}{y^2}\,.
\eeq
One can check that \eq{APR} is valid for $\gamma_0(x,y)$ given by \eq{GA05} at least as far as the 
most singular terms are concerned.

Using \eq{GA05} we can calculate   the ratio
\beq \label{FR}
 \frac{\,\int\,d^2
\,\hat{x}\,d^2\,\hat{y}\,d^2
\tilde{x}\,d^2\,\tilde{y}\,\,\Theta(\hat{x},\hat{y})\,\,\Theta(\tilde{x},\tilde{y})\,\,\gamma^{BA}
\lb \hat{x},\hat{y}; \tilde{x},\tilde{y} \rb}{{\cal F}^2(\infty;[ \gamma_{0,i}])}
\eeq
First we calculate ${\cal F}(\infty;[ \gamma_{0,i}]$ rewriting \eq{GA05} in the  form of contour 
integral
\beq \label{F1}
\gamma_0(x,y)\,\,=\,\frac{1}{16 \pi^2}\,\,\frac{1}{2 \pi i}
\,\int^{a + i \infty}_{a - i \infty}\,\,\frac{d\,h}{h^3}\,\,\lb \frac{x^2}{y^2} \rb^h
\eeq
Using \eq{F1} we can reduce \eq{A232} to the equation
\beq \label{F2}
{\cal F}(\infty;[ \gamma_{0,i}]\,\,=\,\,\frac{\tk}{\kappa}\,\frac{1}{16 }\,\,\frac{1}{2 \pi 
i}\,\int^{a + i \infty}_{a - 
i \infty}\frac{d\,h}{h^3}\,\int^{R^2}\,x^2 \,d x^2\,\int^1_0 \,d\,t\,(t^h + t^{-h})\,\,\,=\,\,
\eeq
$$
\frac{\tk}{\kappa}\,\frac{1}{16}\,\frac{R^4}{2}\,\,\frac{1}{2 \pi
i}\,\int^{a + i \infty}_{a -
i \infty}\frac{d\,h}{h^3}\,\,\frac{1}{ 1 - h^2}\,\,=\,\,\frac{\tk}{\kappa}\,\frac{R^4}{32}
$$
In \eq{F2}  $t$ is the new variable $t = x^2/y^2$.

We use the same trick  to calculate the numerator of \eq{FR}. We, finally,  obtain 
\beq \label{GA08}
 \frac{\,\int\,d^2
\,\hat{x}\,d^2\,\hat{y}\,d^2
\tilde{x}\,d^2\,\tilde{y}\,\,\Theta(\hat{x},\hat{y})\,\,\Theta(\tilde{x},\tilde{y})\,\,\gamma^{BA}
\lb \hat{x},\hat{y}; \tilde{x},\tilde{y} \rb}{{\cal F}^2(\infty;[ \gamma_{0,i}])}\,\,=\,\,\frac{32 
\kappa}{\tk^2}
\eeq

Therefore,  we have
\beq \label{A23SOL1}
N\lb x,y;x',y'; \infty \rb \,\,\,=\,\,\,\,{}_2F_0 \lb  \tk,\tk;\,; - \frac{32\,\kappa}{\tk^2} 
\rb\,\,\,<\,\,\,1\,\,-\,\,e^{- 32\,\kappa}
\eeq

From \eq{A23SOL1}  we see that the amplitude turns out to be less that unity, and the unitarity limit 
cannot be reached. Therefore, we have  gray disc  scattering instead of the black disc one which 
we expect.

\subsubsection{Corrections to the asymptotic solution}
 
Substituting \eq{APA1} into \eq{PHEQ} with $\Gamma(2 \to 1)=0$ we obtain the equation for $\Delta 
\Phi$, namely,
$$
\frac{ \partial\, \Delta \Phi(Y;[\gamma_i])}{\partial\,{\cal Y}}\,\,\,=\,\,-\,\int  \,\,d^2\,x\,d^2\,y
\,d^2\,z\,\,\,K(\x,\y;\z)\,\lb \,\gamma(x,y)\,\,-\,\,\gamma(x,z)\,\gamma(z,y)\,\rb \frac{\de\, \Delta 
\Phi(Y;[\gamma_i])}{\de\,\gamma(x,y)}\,\,+
$$
\beq \label{AP23}
+\,\,2\,\frac{1}{{\cal 
F}(\infty, 
[\gamma_{0i}])}\,\int\,d^2\,z\,d^2\,x_1\,d^2\,y_1\,d^2\,x_2\,d^2\,y_2\,\,\,K(\x_1,\y_2;\z)\,\,\times
\eeq
$$
\lb 
\,\gamma(x_1,y_2)\,\gamma(x_2,y_1)\,\,-\,\,\gamma(x_1,z)\,\gamma(z,y_2)\,\gamma(x_2,y_1)\,\rb 
\,\frac{\de \Delta\Phi(Y;[\gamma_i])}{\de\,\gamma(x_1,y_1)}
$$

This Liouville-type equation  can be  solved   assuming that $\Delta \Phi(Y;[\gamma_i])\,=\,\Delta 
\Phi([\gamma_i(Y;x_i,y_i)])$ as we did in section 4.1.3. For function $\gamma(Y;x,y)$ we obtain the 
equation
$$
\frac{ \partial\,\gamma({\cal Y};x,y)}{\partial\,{\cal Y}}\,\,\,=\,\,-\,\int  \,\,d^2\,x\,d^2\,y
\,\,\,K(\x,\y;\z)\,\lb \,\gamma({\cal Y}; x,y)\,\,-\,\,\gamma({\cal Y};x,z)\,u({\cal Y};z,y)\,\rb \,\,+
$$
\beq \label{AP231}
+\,\,2\,\frac{1}{{\cal
F}}\,\int\,d^2\,z\,d^2\,x_1\,d^2\,y_1\,d^2\,x_2\,d^2\,y_2\,\,\,K(\x_1,\y_2;\z)\,\,\times 
\eeq
$$
\lb
\,\gamma({\cal Y}; x_1,y_2)\,\gamma({\cal Y}; x_2,y_1)\,\,-\,\,\gamma({\cal Y}; x_1,z)\,u({\cal 
Y}; z,y_2)\,\gamma({\cal Y}; x_2,y_1)\,\rb
$$
It is  more convenient to switch to the  funcion $\gamma$: $u({\cal Y};x,y) \,=\,1 
- \gamma({\cal Y};x,y)$.
 Indeed, even from the asymptotic solution we see that the typical values of $u$ should be 
close to unity, namely, $ 1 - u \equiv \gamma \,\approx\,1/\tk$. The relation between functions $u$ 
and/or $\gamma$ is determined by the initial 
condition, namely, $Z({\cal Y}; [u_i])\,=\,u(x,y)$ at $Y =0$. For the functional $\Delta \Phi$, this 
condition can be written 
$$
\Delta \Phi({\cal Y} = 0;[\gamma_i])\,\,=\,\,\ln(1 - \gamma({\cal Y} =0 ; x,y))\,\,+\,\,\tk\,\ln \lb 
\frac{\int\,d^2\,x\,d^2\,y\,\Theta(x,y)\,\,\gamma({\cal Y} =0 ; x,y)}{{\cal F} (\infty; [\gamma_{0,i})} 
\rb\,\,\approx
$$
\beq \label{AP232}
\,\approx\, -\,\gamma({\cal Y} =0 ; 
x,y)\,\,-\,\,\tk\,\ln \frac{\int\,d^2\,x\,d^2\,y\,\Theta(x,y)\,\,\Delta\,\gamma({\cal Y} =0 ; 
x,y)}{{\cal F} 
(\infty; [\gamma_{0,i}])}
\eeq
where $\Delta \gamma({\cal Y};x,y) \,\equiv\,  \gamma({\cal Y};x,y) -  \gamma_0({\cal Y};x,y)$.
The key idea of our approach to corrections at high energies, is to take   $\Delta \gamma 
\,\ll\,\gamma_0$. In the derivation of \eq{AP232} we have used this assumption.

We can also write the relation for $\int\,d^2x\,d^2\,y\,\, \Delta \Phi({\cal Y} = 0;[\gamma_i]) $, 
namely,
\beq \label{AP233}
\int\,d^2x\,d^2\,y\,\, \Delta \Phi({\cal Y} = 0;[\gamma_i])\,\,\,=\,\,{\cal F} ( \infty; 
[\gamma_{0,i}]) \,\,- \,\,\int 
\,d^2\,x\,d^2\,y\,\Theta(x,y)\,\,\Delta\,\gamma({\cal Y} =0 ; x,y)\,\times
\eeq
$$
 \lb \,1\,\,- \,\,\tk\,\frac{N}{
{\cal F} ( \infty;[\gamma_{0,i}])} \rb 
$$
Using \eq{AP232} and \eq{AP233} we obtain that
\beq \label{AP234}
\gamma({\cal Y}=0; x,y)\,\,=\,\,-\,\,\Delta\,\Phi({\cal Y} = 0; x,y)\,\,+\,\,\frac{\tk}{{\cal F}\,(
\infty;[\gamma_{0,i}]) + \tk \,N} \times
\eeq
$$
\,\lb \int\,d^2\,x\,d^2\,y\,\, \Delta \Phi({\cal Y} = 0; x,y)\,+\, {\cal F}( \infty;[\gamma_{0,i}])\rb 
$$

Introducing $ \gamma({\cal Y}; x,y) \,=\, \Delta\,({\cal Y}; x,y) \,+\,\gamma_0(\infty; x,y)$ and 
considering $  \Delta\,({\cal Y}; x,y)\,\ll\,\gamma_0(\infty; x,y)$ we rewrite \eq{AP231} as a linear 
equation with respect to $ \Delta\,({\cal Y}; x,y)$, namely

\beq \label{AP235}
\frac{\partial\,\,\Delta\,\gamma({\cal Y}; x,y)}{\partial\,\,{\cal Y}}\,\,=\,\,
-\,\,\int\,\,d^2\,z \,K(\x,\y;\z)\,\lb \Delta\,\gamma({\cal Y}; x,z) \,+\,\Delta\,\gamma({\cal Y}; 
z,y)\,-\,\Delta\,\gamma({\cal Y}; x,y) \rb
\eeq
 
\eq{AP235} is the BFKL equation,  but with extra sign 
minus in front. Actually, this is the same equation as \eq{ASAFIN} which has been solved (see 
\eq{APSOL1}).   The solution can be written in the form:

\beq \label{AP238} 
\Delta\,\gamma({\cal Y};x,y)\,\,=\,\,\Delta\,\gamma({\cal Y} = 0;  x,y)\,\exp \lb -\,\,\frac{1 - 
\gamma_{cr}}{2\,\,\chi(\gamma_{cr})}\,z^2 \rb
\eeq
where  $z$ is defined in \eq{NVZ}. 

The initial condition of \eq{AP234} determines the relation between function $\gamma$ and the 
functional $\Delta \Phi$. It is even more convenient to use the initial conditions in the 
form of \eq{AP234} or even of \eq{AP232}

Using the unitarity constraint of \eq{TUNG}, and following  the same line of calculation as in the 
derivation of \eq{A23SOL} we obtain:
$$ 
N\lb x,y;x',y',Y\rb \,\,=\,\,N\lb x,y;x',y',\infty\rb\,\,
\lb \,1\,-\,\,(1 + \tk)\times \right.
$$
\beq \label{AP239}
\left.
\lb \exp \lb  -\, \bas\, \frac{1 - \gamma_{cr}}{2\,\,\chi(
\gamma_{cr}))}\,\,(Y -Y')^2 \rb \,\,+\,\,\exp \lb  - \, \bas\,\frac{1 - \gamma_{cr}}{2\,\,\chi(
\gamma_{cr}))}\,\,(Y')^2 \rb \rb \rb
\eeq

We assume that $Y'$ and $Y-Y'$ are much larger than $\ln (x - y)^2 Q^2_s$. Once more the minimal 
corrections occur at $y' = \h Y$ and we have 
\beq \label{AP2310}
R_{\tk}\,\,=\,\,\frac{N\lb x,y;x',y',Y\rb\,
\,\,-\,\,N_{\tk}\lb 
x,y;x',y',\infty\rb}{N_{\tk}\lb
x,y;x',y',\infty\rb}\,\,\propto\,\,\exp \lb\, - \, \bas\,\frac{1 - \gamma_{cr}}{8\,\,\chi(
\gamma_{cr}))}\,\,(Y)^2\, \rb
\eeq
By index $\tk$ we denote the fact that we consider the case with $\Gamma(2 \to 1)=0$.
$N_{\tk}\lb x,y;x',y',\infty\rb $ is given by \eq{A23SOL1}.

In spite of the fact that the asymptotic solutions look quite different, both examples approach 
their 
asymptotic behaviour at the same rate.

\subsubsection{General property of the solution}

\eq{AP2310} shows that the general solution  has the following attractive  features
\begin{itemize}
\item\,\quad The asymptotic solution is not unity,  but a function of coordinates which is smaller 
than 1. This  means that the high energy scattering  corresponds  rather to the gray disc regime
and  not to the black disc one  which was expected;
\item\,\quad The approach to the asymptotic solution is a very fast decreasing  function of energy
 which is proportional to
$ e^{ - C Y^2}$. The coefficient $C$ in 4 times smaller that for the solution \cite{LT}  of the 
Balitsky-Kovchegov 
equation and in 2 times smaller than for the Iancu-Mueller solution \cite{IM}.
\item\,\quad The entire dynamics can be found,  neglecting the process of the merging of two dipoles 
into 
one dipole. This process we need only to determine the value of $\gamma_0$, but it does not affect 
any 
qualitative properties of the asymptotic behaviour of the scattering amplitude at high energy.
\end{itemize}

In general we found the asymptotic solution to the evolution equation (see \eq{ZEQ}) and we prove that this solution is 
selfconsistent. It means, that corrections to this solution decrease with energy. However, we did not prove that our 
solution satisfies the correct initial conditions for the generating functional. Therefore, in principle, we could 
overlook a solution which is diffrent from ours but satisfies the initial condition while ous does not satisfy them. Such 
scenario looks unlikely to us because of our experience with the toy model and because of the exact solution to 
\eq{GENSOL1}. These examples show that we have a correct procedure for finding the asymptotic solution.  

\section{Conclusions}
In this paper we found for the first time,  the high energy amplitude in QCD taking into account the 
Pomeron loops. We found that the processes of merging between two dipoles as well as processes of 
transition of two dipoles into three dipoles crucially change  the high energy asymptotic 
behaviour of 
the scattering amplitude.

Using the fact that in QCD,  $\Gamma(1 \to 2)/\Gamma(2 \to 1)\,\gg\,1$ as well as $ \Gamma(1 \to 
2)/\Gamma(2 \to 3)\,\gg\,1$ we developed a semi-classical method for searching the solution.
  The main results of our approach are presented in \eq{FASN},\eq{ASPFIN1}, \eq{A23SOL1} and
\eq{AP2310}. They show several unexpected features of high scattering in QCD: (i) the asymptotic
amplitude is a function of dipole sizes and, therefore,  the scattering amplitude
describes the gray disc structure at high energy, instead of black disc regime which was expected;
(ii) the solution approaches  the asymptotic in the same way  as the 
solution to the
Balitsky-Kovchegov equation ($\propto\,\exp(- C Y^2) $ ) but coefficient $C$ in four times smaller
than for the Balitsky-Kovchegov equation; (iii) the process of the  merging of two dipoles into one 
only  influences  the initial condition for the generating functional. Indeed, the new initial 
conditions is  $Z(Y; 
[u_i = 1 - 
\gamma_{0,i}]) =1$ instead of  $Z(Y; [u_i = 1]) = 1 $. The value of $\gamma_0$ is determined by 
the  process of the  merging of two dipoles into one, and it turns out to be of the order of 
$\tk/\kappa \,\ll 
\,1$.  Putting this new initial condition we can use
  the Balitsky-JIMWLK approach
\cite{B,JIMWLK} for description of  the high energy  asymptotic behaviour of the scattering 
amplitude without
any modifications recently suggested  \cite{KOLU1}.

One of the experimental consequences that stems from our solution, is the fact that we do not expect
 geometrical scaling behaviour of the amplitude in the saturation region. We also do not expect the 
black disc behaviour, which in the past was  considered as the most plausible for the high energy 
asymptotic
behaviour.

In general,  the generating functional approach has many advantages, in particular, it is very simple  with a clear 
control of  physics in each step of evolution. However, it has its own shortcoming: inside of this approach we cannot 
restrict ourselves by the finite nuimber of possible transitions \footnote{ We thank our referee who rightly noticed 
that we need to discuss this point in the paper.}. For example, why we included $\Gamma(2 \to 3)$ and did not include 
$\Gamma(3 \to 2)$? Formal argument is  that we included all vertices of  the order of $1/N^2_c$ and neglected all 
vertices of the order of $1/N^3_c$. However, without $\Gamma(3 \to 2)$ we do not have symmetric approach and, strictly 
speaking,  we cannot guarantee the $t$-channel unitarity. Since we use $t$-channel unitarity in our estimates for the 
scattering amplitude,  we obtain the answer which respects the $t$-channel unitarity, but, in the spirit of Iancu-Mueller 
factorization \cite{IM}, this answer is correct only for limited range of energy.   The work is in progress for taking 
into account $\Gamma(3 \to 2)$, however, the final answer to the question which vertices we need to include, perhaps, 
lies in more general theoretical approaches (see Refs, \cite{KOLU1,KOLU2,KOLU3,BIT,HIMS,MMSHWO}).

\section*{Acknowledgments:}
We want to thank Asher Gotsman  for very useful
discussions on the subject
of this paper. A special thanks goes to Michael Lublinsky whose sceptical questions encouraged a fresh 
and unbiased approach to the problem in question.

This research was supported in part  by the Israel Science Foundation,
founded by the Israeli Academy of Science and Humanities.


\begin{thebibliography}{99}
\bibitem{GLR}
L. V. Gribov, E. M. Levin and M. G. Ryskin, Phys. Rep. {\bf 100}
(1983) 1.
%%CITATION = PRPLC,100,1;%%

\bibitem{MUQI}
A. H. Mueller and J. Qiu,  Nucl. Phys. {\bf B 268} (1986) 427.
%%CITATION = NUPHA,B268,427;%%
\bibitem{BFKL}
 E.~A.~Kuraev, L.~N.~Lipatov and V.~S.~Fadin,
  %``The Pomeranchuk Singularity In Nonabelian Gauge Theories,''
  Sov.\ Phys.\ JETP {\bf 45} (1977) 199
  [Zh.\ Eksp.\ Teor.\ Fiz.\  {\bf 72} (1977) 377]\,\,;
  %%CITATION = SPHJA,45,199;%%
I.~I.~Balitsky and L.~N.~Lipatov,
  %``The Pomeranchuk Singularity In Quantum Chromodynamics,''
  Sov.\ J.\ Nucl.\ Phys.\  {\bf 28} (1978) 822
  [Yad.\ Fiz.\  {\bf 28} (1978) 1597].
  %%CITATION = SJNCA,28,822;%%
                                                                                                                               

\bibitem{BART}
J.~Bartels, M.~Braun and G.~P.~Vacca,
{\it ``Pomeron vertices in perturbative QCD in diffractive scattering,''}
arXiv:hep-ph/0412218;\,\,\,
M.~Braun,
Eur.\ Phys.\ J.\ C {\bf 16}, 337 (2000)
[arXiv:hep-ph/0001268];\,\,\,
%%CITATION = HEP-PH 0001268;%%
J.~Bartels and C.~Ewerz,
JHEP {\bf 9909}, 026 (1999)
[arXiv:hep-ph/9908454]\,\,\,
%%CITATION = HEP-PH 9908454;%%
M.~Braun,
Eur.\ Phys.\ J.\  {\bf C6}, 321 (1999)
[arXiv:hep-ph/9706373];\,\,\,
%%CITATION = HEP-PH 9706373;%%
%%CITATION = HEP-PH 0412218;%%
M.~A.~Braun and G.~P.~Vacca,
Eur.\ Phys.\ J.\ C {\bf 6}, 147 (1999)
[arXiv:hep-ph/9711486];\,\,\,
%%CITATION = HEP-PH 9711486;%%
J.~Bartels and M.~Wusthoff,
Z.\ Phys.\ C {\bf 66}, 157 (1995).
%%CITATION = ZEPYA,C66,157;%%
\,\,\,\,A.~H.~Mueller and B.~Patel,
Nucl.\ Phys.\ B {\bf 425}, 471 (1994)
[arXiv:hep-ph/9403256];\,\,\,
%%CITATION = HEP-PH 9403256;%%
J.~Bartels,
Z.\ Phys.\  {\bf C60}, 471 (1993);\,\,\,\,
%%CITATION = ZEPYA,C60,471;%%
\bibitem{NP}
H.~Navelet and R.~Peschanski,
Nucl.\ Phys.\,  {\bf B634} (2002) 291
[arXiv:hep-ph/0201285]; \,\,\,
Phys.\ Rev.\ Lett.\  {\bf 82} (1999) 137,,
[arXiv:hep-ph/9809474];\,\,\,
Nucl.\ Phys.\  {\bf B507} (1997)  353,
[arXiv:hep-ph/9703238].



\bibitem{BLV}
J.~Bartels, L.~N.~Lipatov and G.~P.~Vacca,
Nucl.\ Phys.\  {\bf B706}, 391 (2005)
[arXiv:hep-ph/0404110].
%%CITATION = HEP-PH 0404110;%%

\bibitem{GRPO}
P.~Grassberger and K.~Sundermeyer,
%``Reggeon Field Theory And Markov Processes,''
Phys.\ Lett.\  {\bf B77} (1978) 220.
%%CITATION = PHLTA,B77,220;%%
\bibitem{LEPO}
E.~Levin,
%``'Hard' pomeron approach to 'soft' processes at high-energy,''
Phys.\ Rev.\ D {\bf 49} (1994) 4469.
%%CITATION = PHRVA,D49,4469;%%
\bibitem{BOPO}
K.~G.~Boreskov,
{\it ``Probabilistic model of Reggeon field theory,''}
arXiv:hep-ph/0112325 and reference therein.
%%CITATION = HEP-PH 0112325;%%

\bibitem{GARD}
C.W. Gardiner,{\it ``Handbook of Stochastic Methods for Physics, Chemistry
and the
Natural
Science"}, Springer-Verlag, Berlin, Heidelberg 1985.



\bibitem{MUCD}
A.~H.~Mueller, Nucl.\ Phys. {\bf B 415} (1994) 373;
{\it ibid}  {\bf B 437} (1995) 107.





\bibitem{LL}
E.~Levin and M.~Lublinsky,
%``A linear evolution for non-linear dynamics and correlations in  realistic
%nuclei,''
Nucl.\ Phys.\, {\bf A730} (2004) 191
[arXiv:hep-ph/0308279].
%%CITATION = HEP-PH 0308279;%%
\bibitem{LLB}
E.~Levin and M.~Lublinsky,
Phys.\ Lett.\  {\bf B607} (2005) 131
[arXiv:hep-ph/0411121].
%%CITATION = HEP-PH 0411121;%%

\bibitem{B}
I.~Balitsky,
[arXiv:hep-ph/9509348];\,\,
%%CITATION = HEP-PH 9509348;%%
Phys.\ Rev.\ D {\bf 60}, 014020 (1999)
[arXiv:hep-ph/9812311].
%%CITATION = HEP-PH 9812311;%%



\bibitem{K}
Y.~V.~Kovchegov,
%``Small-x F2 structure function of a nucleus including multiple pomeron
%exchanges,''
Phys.\ Rev.\  {\bf D60} (1999) 034008
[arXiv:hep-ph/9901281].
%%CITATION = HEP-PH 9901281;%%


\bibitem{JANIK}
R.~A.~Janik,
{\it ``B-JIMWLK in the dipole sector,''}
arXiv:hep-ph/0409256;\,\,\,\,
%%CITATION = HEP-PH 0409256;%%
R.~A.~Janik and R.~Peschanski,
%``QCD saturation equations including dipole-dipole correlation,''
Phys.\ Rev.\, {\bf D70} (2004) 094005
[arXiv:hep-ph/0407007].
%%CITATION = HEP-PH 0407007;%%



\bibitem{MV}
L. McLerran and R. Venugopalan,  Phys. Rev.  {\bf D 49} (1994)
2233, 3352; {\bf D 50} (1994) 2225, {\bf D 53} (1996) 458, {\bf D 59}
(1999) 09400.

\bibitem{JIMWLK}
~J.~Jalilian-Marian, A.~Kovner, A.~Leonidov and H.~Weigert,
 Phys.\ Rev.\,  {\bf D59} (1999) 014014
[arXiv:hep-ph/9706377];\,\,  Nucl.\ Phys.\,{\bf B504} (1997) 415
[arXiv:hep-ph/9701284];\\ \,\,\,
J.~Jalilian-Marian, A.~Kovner and H.~Weigert,
  Phys.\ Rev.\  {\bf D59} (1999) 014015
  [arXiv:hep-ph/9709432].
  %%CITATION = HEP-PH 9709432;%%
 A.~Kovner, J.~G.~Milhano and H.~Weigert,
  Phys.\ Rev.\ D {\bf 62} (2000) 114005
  [arXiv:hep-ph/0004014]\,; \,\,\,
  %%CITATION = HEP-PH 0004014;%%
E.~Iancu, A.~Leonidov and L.~D.~McLerran,
 Phys.\ Lett.\,  {\bf B510} (2001) 133
[arXiv:hep-ph/0102009];\,\,  Nucl.\ Phys.\,  {\bf A692} (2001) 583
[arXiv:hep-ph/0011241];\\
E.~Ferreiro, E.~Iancu, A.~Leonidov and L.~McLerran,
  Nucl.\ Phys.\  {\bf A703} (2002) 489,
  [arXiv:hep-ph/0109115];\\
  %%CITATION = HEP-PH 0109115;%%
H.~Weigert,
 Nucl.\ Phys.\,  {\bf A703} (2002) 823
[arXiv:hep-ph/0004044].

\bibitem{KOLU1}
A.~Kovner and M.~Lublinsky,
{\it ``In pursuit of pomeron loops: The JIMWLK equation and the Wess-Zumino term,''}
arXiv:hep-ph/0501198.
%%CITATION = HEP-PH 0501198;%%

\bibitem{KOLU2}
A.~Kovner and M.~Lublinsky,
{\it ``Remarks on High Energy Evolution,''}
arXiv:hep-ph/0502071.
%%CITATION = HEP-PH 0502071;%%
\bibitem{KOLU3}
A.~Kovner and M.~Lublinsky,
 {\it ``From target to projectile and back again: Selfduality of high energy
  evolution,''}
  arXiv:hep-ph/0502119.
  %%CITATION = HEP-PH 0502119;%%
\bibitem{BIT}
J.~P.~Blaizot, E.~Iancu, K.~Itakura and D.~N.~Triantafyllopoulos,
 {\it ``Duality and Pomeron effective theory for QCD at high energy and large
 $ N_c$,''}
  arXiv:hep-ph/0502221.
  %%CITATION = HEP-PH 0502221;%%

\bibitem{IM}
E.~Iancu and A.~H.~Mueller,
 Nucl.\ Phys.\,  {\bf A730} (2004) 460, 494,
[arXiv:hep-ph/0308315],[arXiv:hep-ph/0309276].
%%CITATION = HEP-PH 0309276;%%
%%CITATION = HEP-PH 0308315;%%


\bibitem{KOLE}
M.~Kozlov and E.~Levin,
%``The Iancu-Mueller factorization and high energy asymptotic behaviour,''
Nucl.\ Phys.\  {\bf A739} (2004) 291
[arXiv:hep-ph/0401118].
%%CITATION = HEP-PH 0401118;%%

\bibitem{MUSO}
A.~H.~Mueller and A.~I.~Shoshi,
%``Small-x physics beyond the Kovchegov equation,''
Nucl.\ Phys.\ B {\bf 692} (2004) 175
[arXiv:hep-ph/0402193].
%%CITATION = HEP-PH 0402193;%%


\bibitem{IT1}
E.~Iancu and D.~N.~Triantafyllopoulos,
{\it ``A Langevin equation for high energy evolution with pomeron loops,''}
arXiv:hep-ph/0411405.
%%CITATION = HEP-PH 0411405;%%

\bibitem{MSW}
A.~H.~Mueller, A.~I.~Shoshi and S.~M.~H.~Wong,
{\it ``Extension of the JIMWLK equation in the low gluon density region,''}
arXiv:hep-ph/0501088.
%%CITATION = HEP-PH 0501088;%%

\bibitem{LLTR}
E.~Levin and M.~Lublinsky,
{\it ``Towards a symmetric approach to high energy evolution: Generating functional
with Pomeron loops,''}
arXiv:hep-ph/0501173.
%%CITATION = HEP-PH 0501173;%%




\bibitem{IT2}
E.~Iancu and D.~N.~Triantafyllopoulos,
{\it ``Non-linear QCD evolution with improved triple-pomeron vertices,''}
arXiv:hep-ph/0501193.
%%CITATION = HEP-PH 0501193;%%

\bibitem{MUCH}
Z.~Chen and A.~H.~Mueller,
%``The Dipole picture of high-energy scattering, the BFKL equation and many
%gluon compound states,''
Nucl.\ Phys.\  {\bf B451} (1995) 579.
%%CITATION = NUPHA,B451,579;%%

\bibitem{HIIM}
Y.~Hatta, E.~Iancu, K.~Itakura and L.~McLerran,
{\it``Odderon in the color glass condensate,''}
arXiv:hep-ph/0501171.
%%CITATION = HEP-PH 0501171;%

\bibitem{VE}
~G. ~Veneziano, {\it Phys. Letters} {\bf 52B} (1974) 220; {\it Nucl.
Phys.}{\bf B74} (1974) 365;\\
~M. ~Ciafaloni, ~G. ~Marchesini anf ~G. ~Veneziano,
{\it Nucl. Phys.} {\bf B98} (1975) 493.




\bibitem{BKP}
J.~Bartels,
% ``High-Energy Behavior In A Nonabelian Gauge Theory. 2. First
%Corrections To
%T(N--->M) Beyond The Leading Lns Approximation,''
%
Nucl.\ Phys.\  {\bf B175}, 365 (1980);\,\,\,\,
%%CITATION = NUPHA,B175,365;%%
J.~Kwiecinski and M.~Praszalowicz,
% ``Three Gluon Integral Equation And Odd C Singlet Regge Singularities In
%QCD,''
%
Phys.\ Lett.\  {\bf B94}, 413 (1980).
%%CITATION = PHLTA,B94,413;%%

\bibitem{LI}
L.~N.~Lipatov,
Phys.\ Rept.\  {\bf 286}, 131 (1997)
[arXiv:hep-ph/9610276];\,\,Sov. Phys. JETP {\bf 63} (1986) 904 and 
references therein.
\bibitem{KKM}
G.~P.~Korchemsky, J.~Kotanski and A.~N.~Manashov,
Phys.\ Lett.\  {\bf B583} (2004) 121
[arXiv:hep-ph/0306250] \,\,\,;
S.~E.~Derkachov, G.~P.~Korchemsky, J.~Kotanski and A.~N.~Manashov,
Nucl.\ Phys.\  {\bf B645} (2002) 237
[arXiv:hep-th/0204124];\,\,\,
%%CITATION = HEP-TH 0204124;%%
S.~E.~Derkachov, G.~P.~Korchemsky and A.~N.~Manashov,
Nucl.\ Phys.\  {\bf B661} (2003) 533
[arXiv:hep-th/0212169];\,\,\,
%%CITATION = HEP-TH 0212169;
Nucl.\ Phys.\  {\bf B617} (2001) 375
[arXiv:hep-th/0107193];\,\,\,
%%CITATION = HEP-TH 0107193;%%
and references therein.
%%CITATION = HEP-PH 0306250;%%




\bibitem{WE}
H.~Weigert,
{\it ``Evolution at small $x_{bj}$: The color glass condensate,''}
arXiv:hep-ph/0501087 and references therein.
%%CITATION = HEP-PH 0501087;%%
\bibitem{BIW}
J.~P.~Blaizot, E.~Iancu and H.~Weigert,
%``Non linear gluon evolution in path-integral form,''
Nucl.\ Phys.\  {\bf A713} (2003) 441
[arXiv:hep-ph/0206279].
%%CITATION = HEP-PH 0206279;%%


\bibitem{BAA}
I. F. Ginzburg, S.L. Panfil and V.G. Serbo, Nucl. Phys. {\bf B284} (1987)
685, {\bf B296} (1988) 569;\,\,I. F. Ginzburg and D. Yu. Ivanov,  Nucl.
Phys. {\bf B388} (1992) 376;\,\,\,D.~Y.~Ivanov and R.~Kirschner,
Phys.\, Rev.\,  {\bf D58} (1998) 114026,
{\tt hep-ph/9807324};\,\,M.~Kozlov and E.~Levin,
%``QCD saturation and gamma* gamma* scattering,''
Eur.\ Phys.\ J.\ C {\bf 28} (2003) 483
[arXiv:hep-ph/0211348].
%%CITATION = HEP-PH 0211348;%%


\bibitem{RY}
I. Gradstein and I. Ryzhik, {\it `` Tables of Series, Products, and
Integrals"}, Verlag MIR, Moskau,1981.

\bibitem{GS}
~J.~Kwiecinski and A.~M.~Stasto,
 Acta Phys.\ Polon.\, {\bf B33} (2002) 3439;\,\,{\it Phys.\ Rev.}\,
{\bf
D66}
(2002) 014013
[arXiv:hep-ph/0203030];\,\,\,\,\,A.~M.~Stasto, K.~Golec-Biernat and
J.~Kwiecinski,
 Phys.\ Rev.\ Lett.\,  {\bf 86} (2001) 596
arXiv:hep-ph/0007192];\,\,\,
J.~Bartels and E.~Levin,
 Nucl.\ Phys.  {\bf B387} (1992) 617;\,\,\,E.~Iancu, K.~Itakura and
L.~McLerran,
 Nucl.\ Phys. \, {\bf A708} (2002) 327
[arXiv:hep-ph/0203137].
%%CITATION = APPOA,B33,3439;%%
%%CITATION = HEP-PH 0007192;%%
%%CITATION = NUPHA,B387,617;%%
%%CITATION = HEP-PH 0203137;%%


\bibitem{BOKOLE}
S.~Bondarenko, M.~Kozlov and E.~Levin,
%``QCD saturation in the semi-classical approach,''
Nucl.\ Phys.\  {\bf A727} (2003) 139
[arXiv:hep-ph/0305150].
%%CITATION = HEP-PH 0305150;%%


\bibitem{MP}
S.~Munier and R.~Peschanski,
Phys.\ Rev.\  {\bf D70} (2004) 077503;
 {\bf D69} (2004) 034008
[arXiv:hep-ph/0310357];\,\,
{\it Phys.\ Rev.\ Lett.}\,  {\bf 91} (2003) 232001
[arXiv:hep-ph/0309177].

\bibitem{FROI}
M.~Froissart,
%``Asymptotic Behavior And Subtractions In The Mandelstam Representation,''
Phys.\ Rev.\  {\bf 123}, 1053 (1961).
%%CITATION = PHRVA,123,1053;%%
\bibitem{FRDIS}
A.~L.~Ayala, M.~B.~Gay Ducati and E.~M.~Levin,
%``Unitarity boundary for deep inelastic structure functions,''
Phys.\ Lett.\ B {\bf 388} (1996) 188
[arXiv:hep-ph/9607210].
%%CITATION = HEP-PH 9607210;%%
\bibitem{IMM}
 E.~Iancu, A.~H.~Mueller and S.~Munier,
  Phys.\ Lett.\  {\bf B606} (2005) 342
  [arXiv:hep-ph/0410018].
  %%CITATION = HEP-PH 0410018;%%


\bibitem{LT}
E.~Levin and K.~Tuchin,
%``Nonlinear evolution and saturation for heavy nuclei in DIS,''
Nucl.\ Phys.\  {\bf A693} (2001) 787
[arXiv:hep-ph/0101275];\,\,\,{\bf A691} (2001) 779
[arXiv:hep-ph/0012167];\,\,\,{\bf B573} (2000) 833
[arXiv:hep-ph/9908317].
%%CITATION = HEP-PH 9908317;%%

%%CITATION = HEP-PH 0012167;%%
%%CITATION = HEP-PH 0101275;%%

\bibitem{KLBK}
 M. ~Kozlov and E.~Levin, in preparation.
\bibitem{MU02}
A.~H.~Mueller, {\it Nucl.\ Phys.}\,  {\bf B643} (2002) 501
[arXiv:hep-ph/0206216]

\bibitem{MUTR}
A.~H.~Mueller and D.~N.~Triantafyllopoulos,
%``The energy dependence of the saturation momentum,''
Nucl.\ Phys.\  {\bf B640} (2002) 331
[arXiv:hep-ph/0205167];\,\,\,\,D.~N.~Triantafyllopoulos,
%``The energy dependence of the saturation momentum from RG improved BFKL
%evolution,''
Nucl.\ Phys.\ B {\bf 648} (2003) 293
[arXiv:hep-ph/0209121].
%%CITATION = HEP-PH 0209121;%%
%%CITATION = HEP-PH 0205167;%%

\bibitem{Kancheli}
  O.~V.~Kancheli,
  {\it ``About the structure of the Froissart limit in QCD,''}
  arXiv:hep-ph/0008299.
  %%CITATION = HEP-PH 0008299;%%
\bibitem{FR}
E.~Ferreiro, E.~Iancu, K.~Itakura and L.~McLerran,
  %``Froissart bound from gluon saturation,''
  Nucl.\ Phys.\ A {\bf 710} (2002) 373
  [arXiv:hep-ph/0206241]\,\,;
  %%CITATION = HEP-PH 0206241;%%
 E.~M.~Levin and M.~G.~Ryskin,
  %``High-Energy Hadron Collisions In QCD,''
  Phys.\ Rept.\  {\bf 189} (1990) 267.
  %%CITATION = PRPLC,189,267;%%

\bibitem{CRPOM}
  A.~A.~Migdal, A.~M.~Polyakov and K.~A.~Ter-Martirosian,
  %``Theory Of Interacting Pomerons,''
  Phys.\ Lett.\ B {\bf 48} (1974) 239
  [Pisma Zh.\ Eksp.\ Teor.\ Fiz.\  {\bf 68} (1975) 817].
  %%CITATION = PHLTA,B48,239;%%
\bibitem{HIMS}
Y.~Hatta, E.~Iancu, L.~McLerran and A.~Stasto,
  {\it ``Color dipoles from bremsstrahlung in QCD evolution at high 
energy,''}
  arXiv:hep-ph/0505235\,\,;
  %%CITATION = HEP-PH 0505235;%%
Y.~Hatta, E.~Iancu, L.~McLerran, A.~Stasto and D.~N.~Triantafyllopoulos,
  {\it ``Effective Hamiltonian for QCD evolution at high energy,}
  arXiv:hep-ph/0504182.
  %%CITATION = HEP-PH 0504182;%%
\bibitem{MMSHWO}
  C.~Marquet, A.~H.~Mueller, A.~I.~Shoshi and S.~M.~H.~Wong,
  {\it ``On the projectile-target duality of the color glass condensate in 
the
  dipole picture,''}
  arXiv:hep-ph/0505229.
  %%CITATION = HEP-PH 0505229;%%




\end{thebibliography}
\end{document}